%% file: fastmat.tex
\documentclass[12pt, a4paper]{elsarticle}

\usepackage{amssymb}
\usepackage{amsmath}
\usepackage{amsthm}
\usepackage{bm}
\usepackage{float}
\usepackage{lineno}
\usepackage[frozencache]{minted}
\usepackage{placeins}
\usepackage{setspace}

\usepackage{tikz,pgfplots}
\pgfplotsset{compat = newest}
\usetikzlibrary{shapes.misc, positioning}
\usetikzlibrary{fit}
\usetikzlibrary{positioning, snakes, decorations}
\usetikzlibrary{patterns}
\usetikzlibrary{shapes.arrows, fit, positioning, calc, decorations}
\usetikzlibrary{external}
\usetikzlibrary[pgfplots.colormaps]
\pgfplotsset{compat=newest,}
\usepgfplotslibrary{groupplots}
\usepgfplotslibrary{dateplot}

\DeclareMathOperator{\FFT}{fft}
\DeclareMathOperator{\iFFT}{ifft}
\DeclareMathOperator{\Diag}{diag}
\DeclareMathOperator{\BlkDiag}{blkdiag}
\DeclareMathOperator{\Min}{min}

\newcommand{\rev}[1]{{\color{blue}#1}}

\usepackage{lineno}
\usepackage{float}
\restylefloat{table}

\usepackage[
  round-mode=places,
  round-precision=2,
  binary-units=true,
  per-mode=symbol,
]{siunitx}
\DeclareSIUnit\irf{IRF}
\DeclareSIUnit{\sample}{S}
\usepackage{hyperref}
\usepackage{cleveref}

\usepackage{subcaption}

\usepackage{tabu}
\usepackage{booktabs}

\usepackage{glossaries}
\newacronym{api}{API}{Application Programming Interface}
\newacronym{cs}{CS}{Compressed Sensing}
\newacronym{dft}{DFT}{Discrete Fourier Transform}
\newacronym{ista}{ISTA}{Iterative Shrinkage-Thresholding Algorithm}
\newacronym{fista}{FISTA}{Fast Iterative Shrinkage-Thresholding Algorithm}
\newacronym{gp}{GP}{Gaussian Processes}
\newacronym{gnss}{GNSS}{Global Navigational Satellite Systems}
\newacronym{lfsr}{LFSR}{Linear Feedback Shift Register}
\newacronym[longplural={Fast Fourier Transforms}]{fft}{FFT}{Fast Fourier Transform}
\newacronym{omp}{OMP}{Orthogonal Matching Pursuit}
\newacronym{ota}{OTA}{Over-The-Air}
\newacronym{prn}{PRN}{Pseudo-Random Noise}
\newacronym{saft}{SAFT}{Synthetic Aperture Focusing Technique}
\newacronym{ski}{SKI}{Structured Kernel Interpolation}
\newacronym{ssr}{SSR}{Sparse Signal Recovery}
\newacronym{wfs}{WFS}{Wave-Field Synthesis}

\newcommand\fastmat{\texttt{fastmat}~}
\newcommand\Matrix{\texttt{Matrix}~}
\newcommand\Algorithm{\texttt{Algorithm}~}

\newcolumntype{T}{>{\raggedleft\arraybackslash\ttfamily%
}l}

\begin{document}

\begin{frontmatter}



\title{fastmat: Efficient Linear Transforms in Python}


\author{Christoph Wilfried Wagner, Sebastian Semper, Jan Kirchhof}

\address{Electronic Measurements and Signal Processing (EMS) Group,\\Technische Universität Ilmenau, Germany}

\begin{abstract}
Scientific computing requires handling large linear models, which are often composed of structured matrices.
With increasing model size, dense representations quickly become infeasible to compute or store.
Matrix-free implementations are suited to mitigate this problem but usually complicate research and development effort by months, when applied to practical research problems.

Fastmat is a framework for handling large composed or structured matrices by offering an easy-to-use abstraction model.
It allows expressing and using linear operators in a mathematically intuitive way, while maintaining a strong focus on efficient computation and memory storage.
The implemented user interface allows for very readable code implementation with very close relationship to the actual mathematical notation of a given problem.
Further it provides means for quickly testing new implementations and also allows for run-time execution path optimization.

Summarizing, fastmat provides a flexible and extensible framework for handling matrix-free linear structured operators efficiently, while being intuitive and generating easy-to-reuse results.
\\
\end{abstract}


\end{frontmatter}
\nolinenumbers
\section*{Required Metadata}\label{metadata}

\section*{Current code version}\label{current_code}

\begin{table}[H]
\begin{tabular}{|l|p{6.5cm}|p{6.5cm}|}
\hline
\textbf{Nr.} & \textbf{Code metadata description} & \textbf{Please fill in this column} \\
\hline
C1 & Current code version & v0.2 \\
\hline
C2 & Permanent link to code/repository used for this code version & \url{https://github.com/EMS-TU-Ilmenau/fastmat} \\
\hline
C3 & Code Ocean compute capsule & \\
\hline
C4 & Legal Code License   & Apache License Version 2.0 \\
\hline
C5 & Code versioning system used & git \\
\hline
C6 & Software code languages, tools, and services used & Python, Cython \\
\hline
C7 & Compilation requirements, operating environments \& dependencies & \\
\hline
C8 & If available Link to developer documentation/manual & \url{https://fastmat.readthedocs.io/en/latest/} \\
\hline
C9 & Support email for questions & {\small \url{sebastian.semper.tu-ilmenau.de} \url{christoph.wagner@tu-ilmenau.de}}\\
\hline
\end{tabular}
\caption{Code metadata}
\label{mand_metadata}
\end{table}


\section{Motivation and significance}\label{motivation}
\input{sections/motivation.tex}

\section{Software description}\label{description}
\input{sections/description.tex}

\section{Illustrative Examples}\label{examples}
\input{sections/examples.tex}

\section{Impact}\label{impact}
\input{sections/impact.tex}

\section{Conclusions}\label{conclusion}
\input{sections/conclusion.tex}

\section{Conflict of Interest}
We confirm that there are no known conflicts of interest associated with this publication and there has been no significant financial support for this work that could have influenced its outcome.

\section*{Acknowledgments}\label{ack}

We would like to express thanks for the support received during the development of \fastmat by the Carl-Zeiss Foundation under the project ``PRIME'', by the DFG under the projects ``CoSMoS'' and ``HoPaDyn'' and also by the Free State of Thuringia and the European Social Fund through the project HYLOC ``2019FGR0100''.
We acknowledge support for the publication costs by the Open Access Publication Fund of the Technische Universität Ilmenau.




\bibliographystyle{elsarticle-num}
\bibliography{bib.bib}

%
%

\end{document}

%% file: sections/motivation.tex

%
%
%
%

Linear transformations are one of the corner stones in applied math, physics, engineering and data science.
This is due to the fact that often one models objects of interest to be resided in a vector space.
In case of finite dimensional vector spaces, linear mappings are represented by matrices.
A matrix encodes a linear mapping, since it tells us how the images of all basis vectors of a given vector space map can be expressed in the coordinates of the image space.
And due to the linearity of the mappings and the spaces themselves this completely characterizes a specific linear mapping.

However, in many applications the linear mappings one studies or uses are not completely arbitrary, but follow a specific \rev{structure}.
These models often stem from physical models~\cite{knabner2003pdes,Stan1993radon}, descriptions of natural processes~\cite{Ludwig1966radontrafo,Quinto2005Xray_tomo} or structural assumptions~\cite{timothy2011sparsematrices,Shepp1974fourierreco} one puts as side constraints.
One prominent example (and certainly one of the most important) is the \gls{dft}, which transforms periodic discrete signals into the respective frequency domain.
As such it has a plethora of applications in spectral analysis, radar, array processing and beyond.
Given the canonical standard basis in $\mathbb{C}^n$, the corresponding matrix elements are expressed as
\begin{equation}\label{dft_matrix}
	\bm F = \left[f_{i,\rev{k}}\right]_{i,\rev{k} = 1}^{n}
	= \left[\exp\left(
	\frac{-\jmath 2 \pi}{n} \cdot i \cdot \rev{k}
	\right)\right]_{i,\rev{k} = 1}^{n}.
\end{equation}
As we can see, the matrix $\bm F \in \mathbb{C}^{n \times n}$ is highly structured, essentially only needing the integer value $n \in \mathbb{N}$ in order to define it completely.
As such the \gls{dft} matrix has no degrees of freedom, since the size of the involved vector space already fully defines its elements.

In order to actually carry out a linear mapping $\bm m$, which means evaluating it at a given input vector $\bm x$ to get its image $\bm y = \bm m(\bm x)$, one simply calculates $\bm y = \bm M \mkern-2.5mu \cdot \bm x$, if $\bm M \in \mathbb{C}^{m \times n}$ is the matrix encoding the linear mapping $\bm m$.
Generally, this is accomplished by using
\begin{equation}\label{matrix_mult}
	\bm y
	= [y_i]_{i = 1}^m
	= \left[\sum\limits_{j = 1}^\rev{n} m_{i,j} \cdot x_j\right]_{i = 1}^m
	= \bm M \mkern-2.5mu \cdot \bm x.
\end{equation}
Note that this formula entails a computational complexity of $\mathcal{O}(n \cdot m)$, both in terms of runtime and memory consumption.
Coming back to our example of the \gls{dft}, it is well known that one should in fact not use \eqref{matrix_mult} in order to multiply the matrix in \eqref{dft_matrix} to a vector.
Instead, one should make use of the \gls{fft}\rev{~\cite{JohnsonFr08:burrus}} in order to calculate $\bm y = \bm F \mkern-1.5mu \cdot \bm x$ with runtime complexity $\mathcal{O}(n \log n)$. \rev{This algorithm exploits the fact that the Fourier matrix can be decomposed into smaller matrices by a so-called butterfly-structure, which in turn depends on the prime factors of the transform size.}
Its runtime complexity is one of the reasons why modern signal processing is possible and practically feasible and it is due to this that the \gls{dft} is one of the most fundamental transforms.

As illuminated by above example, in many cases the matrix-vector product can be implemented by an efficient algorithm, which is tailored to the specific linear mapping at hand.
This results in \emph{two} representations of a linear mapping. First, we have the matrix as a rectangle of scalars, which we call the \emph{dense} representation.
On the other hand, we have the algorithm that implements the matrix-vector product directly, without the matrix.
Hence we call it the \emph{matrix-free} representation of the mapping.
In case of the \gls{dft} we have the \gls{dft}-matrix $\bm F$ as the dense and the \gls{fft} as the matrix-free representation respectively.
In order to stress the matrix-free representation at some places, we use the \emph{forward transform} $\bm \phi_{\rev{\bm M}} : \mathbb{C}^n \rightarrow \mathbb{C}^n$ as
\begin{equation}\label{forward_trafo}
	\bm x \mapsto \bm \phi_{\bm M}(\bm x) = \bm m(\bm x) = \bm M \bm x
\end{equation}
and the \emph{backward transform }$\bm \beta_{\rev{\bm M}} : \mathbb{C}^n \rightarrow \mathbb{C}^n$ as
\begin{equation}\label{backward_trafo}
	\bm x \mapsto \bm \beta_{\bm M}(\bm x) = \bm M^H \bm x,
\end{equation}
where $\bm M^H$ denotes the Hermitian transpose of the matrix $\bm M$.

In order to illustrate the relationship between these two representations, we consider a \rev{researcher, who has to make use of various linear mappings in order to test prototypes of software or just algorithm snippets.}
In this case the involved mappings are just means to an end and not the focus of research.
However, due to the complexity of the simulations in terms of problem size or desired accuracy of the results, the computational demand is so high that the involved linear mappings have to be represented matrix-free.
This way, bottlenecks in runtime and memory usage can be overcome.
Now the problem is that the matrix-free representation is much harder to handle for a researcher who is merely using it as a tool.
In case of the ``simple'' \gls{fft}, dedicated and highly optimized libraries \cite{FFTW05, JohnsonFr08:burrus, FFTWgen99, FFTW98} exist due to the fact that it is far from trivial to devise numerically stable algorithms for many matrix-free implementations.

Additionally, our conceptual engineer (let's call him Harold) more often than not encounters the algorithms in his field being described in the dense format.
In contrast, the actual implementation is usually matrix-free, to allow for efficient runtime performance -- or sometimes even to make the described algorithm practically feasible in the first place.
However, algebraic notation certainly is easier to digest and manipulate than specifically fine-tuned computer code -- and therefore more convenient and reliable to Harold.
As a consequence, we often see the familiar discrepancy between mathematical descriptions in a publication and its actual implementation counterpart in software.

Let us consider the simple case of fast cyclic convolution with a vector $\bm c \in \mathbb{C}^n$.
It is a linear transformation and as such can be described via $\bm y = \bm C(\bm c) \cdot \bm x$.
However, the actual implementation usually would compute
\begin{equation}\label{fft_convolution}
	\bm y = \iFFT(\FFT(\bm c) \odot \FFT(\bm x)),
\end{equation}
where $\odot: \mathbb{C}^n \times \mathbb{C}^n \rightarrow \mathbb{C}^n$ carries out a pointwise multiplication \rev{(i.e. Hadamard product)} of two vectors in $\mathbb{C}^n$.

We can identify several implications following from the above:
First, the two representations (of the same thing) can become arbitrarily distinct from each other, which makes it harder to cope with them on a theoretical level.
Secondly, the programming code (used to supply one's own work with numerical evidence) diverges from the notation in the publication.
Both facts contribute to a higher risk of faulty implementations, hard to digest implementations and difficult to maintain code bases.

The software we are presenting in this publication aims to provide a programming interface that brings both described representations closer together.
This is realized by providing a \rev{collection} of matrices (see \Cref{zoo_of_matrices}), that can be treated like the dense representation, but internally use the matrix-free representation.
In addition, we provide binary and unary operators and more complicated construction routines to combine these operators to build even more sophisticated structures for linear mappings.
Further, operators on linear mappings themselves may be represented in a matrix-free fashion, too.
Prominent examples are matrix-matrix products, unfoldings of structured tensors and various types of block matrices.
Finally, the basic \gls{api}, which is described in the next section, is kept simple enough so that it poses a low threshold to implement additional linear mappings, \rev{if required}.

%% file: sections/description.tex


\subsection{Software Architecture}\label{architecture}

During the development of the described package we aimed at addressing the problems scientists will ultimately face when working with large or complex linear transforms.
In the following, we first describe these obstacles and derive hence necessary properties of our architecture.
Then, we outline how we implemented these in our software.

Once our prototypical engineer Harold leaves the zone of theory, he has to come up with an implementation of his proposed algorithm in order to present some numerical evidence for his research.
Although Harold is no expert in numerics and programming, he strives for the most efficient implementation in order to experience the validity of his numerical experiments within his lifetime and available resources.
But at the same time he must minimize the effort for arriving at such an implementation, to avoid the trap of simply trading excess runtime against excess development time.
Hence, to avoid these struggles we have to identify their origins, which (as we will see) are plentiful.

Most obviously, the computing resources are not sufficient to carry out a certain part of a na\"ive implementation in a straightforward fashion.
So circumventing bottlenecks in computing hardware is one obstacle Harold might face.
Also, the algorithms he is prototyping are prone to change frequently due to new ideas, new requirements or just theoretical errors during conceptualization.
It can easily happen that these seemingly small changes and fixes force Harold to reorganize and rework large parts of his codebase, which in his case increases the risk for new bugs and and hence frustration.
Once Harold is satisfied with his implementation there is still the researcher's obligation to ensure correctness of his simulations, because otherwise the conclusions he draws from them are debatable at best.
This debugging is something Harold is not proficient in and hence it puts a lot of unwanted and possibly avoidable strain on him.
Finally, not only high-level implementational errors cause trouble for Harold, since culprits may also wait at the lowest levels, such as floating point accumulation errors or other accuracy issues.
It takes a lot of experience to correctly anticipate and counter these in numerical stability analysis, which Harold simply does not have.

Hence we have to ask the question: ``How does our architecture help to Hide the Pain from Harold, when working with linear transforms?''
First, Harold must be enabled to reuse his linear transform-of-choice implementation in any other context, once it succeeded.
Hence we aim at a high level of flexibility and modularity when designing the architecture.
Second, the software itself has to provide extensive, reliable and easily accessible means for testing, such that Harold has the opportunity to test thoroughly early-on during development, when fixes still are not complex -- and hence not expensive.
And finally, the architecture must enable Harold to write code that is as clean, readable and mappable to the algebraic notation as possible.
This especially helps when fellow researchers wish to extend his work or study it more closely.

For this task we have identified the Python ecosystem to be the best choice due to its design philosophy.
Additionally, the availability of Numpy and Scipy allows a good compatibility to other scientific computing packages for Python, since we heavily rely on the \texttt{ndarray} data structure provided by Numpy~\cite{2020NumPy-Array}.
%

\subsubsection{General Matrix-free Representations}\label{general_baseclass}

As we have seen, every linear transform can be represented by a matrix.
Hence, the interaction and usage of linear transforms can always be recast as an interaction and usage of matrices.
In order to provide a general and structure independent implementation for these modifications, properties and interactions, we offer the \Matrix baseclass.

One important feature is that the user must be able to easily extend this baseclass without having to re-implement all of its functionality, while still retaining compatibility with the rest of the package.
In that spirit, the \Matrix baseclass does not make any assumptions about the structure of the linear transform, introducing such an abstract object-oriented model for matrix-free linear operators.
An extensive user interface is offered for reflecting additional structural assumptions in user classes (which by default inherit all general methods from the \Matrix baseclass), just by making a few targeted modifications through method overloading.

A good example for this is the \texttt{Toeplitz} \rev{matrix}, which represents a \rev{non-circular} convolution.
As such it has an efficient forward transform by doing the convolution in the frequency domain and can be inverted quickly by means of Levinson recursion and hence these methods should be implemented differently than in the \Matrix baseclass.
However, when eigenvalues are of interest, these can in general not be calculated explicitly more efficiently than with a general diagonalization method.
As such, inheriting and overloading methods from the \Matrix baseclass describes the path from no structural assumptions of a linear mapping to more constraints, less generality and hence more efficiency.

One thing to note when delving into matrix-free representations is that many quantities of a given matrix $\bm M$ can also be calculated using the forward and backward transforms in \cref{forward_trafo,backward_trafo}.
For instance, the single matrix element $m_{i,j}$ can be accessed via $\bm e_i^T \bm \phi_{\bm M}(\bm e_j)$ independently of the algorithm used for $\bm \phi_{\bm M}$.
Since the baseclass provides means of accessing these quantities in this general fashion, overloading efficient means for $\bm \phi$ and $\bm \beta$ is mostly sufficient to use all features of the architecture with an efficiency boost from underlying structural assumptions.

\begin{table}[t]
  \centering
  \input{figures/figure-baseclass}
  \caption{The \Matrix baseclass: properties, methods, overloading and caching}
  \label{baseclass_features}
\end{table}

An overview of these features can be found in \Cref{baseclass_features}, which also illuminates another important aspect of the user interface.
Properties are used to retrieve static information about a matrix in a general fashion.
Methods on the other hand, return certain calculation results after performing an operation on or with the \Matrix instance.
All of these methods and properties have predefined hooks where a user can override the crucial computation parts, if the structure allows for a more efficient implementation.
For instance, coming back to the example of the \texttt{Toeplitz} matrix, the calculation of \texttt{Toeplitz.getItem($\bullet$)} can be executed without relying on $\bm \phi$ or $\bm \beta$ and should instead be calculated with smart indexing arithmetic.

Another aspect of the general baseclass is the fact that it forces the defined matrix instances to be immutable after initialization.
This is due to a limitation in the Python language that makes it impossible to guarantee that dependent objects (such as pre-computations upon initialization) are updated once their underlying dependencies get modified.

In fact, in context of in-place modification of numerical arrays, this statement even holds true for most hardware computer architectures.
The architectural decision to establish instance immutability is forced as a consequence of the fact, that almost all efficient $\bm \phi$ and $\bm \beta$ implementations greatly benefit from such pre-computations and caching.

\begin{figure}[htp]
  \input{figures/figure-matrices.tex}
  \caption{Listing of fastmat classes by elementary and composite classes}
  \label{zoo_of_matrices}
\end{figure}

A good example is for instance the implementation of the \texttt{Circulant} matrix, which realizes the computation given in \eqref{fft_convolution}. Since $\bm c$ accurately defines the circulant matrix it is beneficial to pre-compute $\FFT(\bm c)$ once and then store it for later calls of $\bm \phi$ and $\bm \beta$.
Instead, the architecture encourages the initialization of a new instance with modified parameters and strongly dis-encourages in-place modification of any \Matrix instance attributes, to avoid consistency issues.

To maximize reusability, the \Matrix instance behaves identical in any context -- regardless of customization -- as long as its use makes sense mathematically.
This incentivizes the definition of new classes by recombining already existing classes, reducing complexity by stacking multiple layers of structural abstraction.

As a manifestation of this philosophy, the \texttt{Circulant} matrix is not implemented by hacking down \eqref{fft_convolution} in closed form, but rather by defining it as a product of three matrices:
A conjugated and transposed Fourier matrix, a diagonal matrix and another Fourier matrix.
Given that we have implemented  the \texttt{Product}, \texttt{Fourier} and \texttt{Diagonal} classes correctly, we can rely on them to serve as abstraction layers for the definition of \texttt{Circulant} with similar efficiency as a closed-form implementation.
See \Cref{zoo_of_matrices} for an overview of currently available types of operators.

Summarizing, for certain types of composite operators, their respective forward and backward transforms can themselves be described solely from the respective transforms of their operator terms.
Structural knowledge is fully retained even with multiple layers of structural abstraction -- meeting the architectural goal of maintaining a close relationship between fastmat-powered code and the theoretical computation description using mathematical notation.


\subsubsection{Matrix-free Algorithm Representation}\label{general_baseclass}

On top of the pure matrices we also offer a \Algorithm baseclass that facilitates the usage of \Matrix instances for more complex routines that only make use of linear transforms.
As such, this class makes it easier to implement algorithms that rely on the class-\gls{api} provided by the transform baseclass.
This results in some significant advantages:
First, the algorithm implementation itself is completely agnostic with respect to the actual structure of the underlying matrices it makes use of.
As a direct consequence, any \Matrix instance, that exploits structural knowledge for efficiency, passes its benefits on to any \Algorithm that uses it.
Secondly, the \Algorithm baseclass offers means to easily establish user callbacks for various purposes.
This way it is possible to easily inspect, tune or override certain aspects of an algorithm during runtime, such as break conditions, step sizes, thresholding or the custom determination of regularization parameters.
And last but not least, the algorithm's code structure can also reflect the mathematical notation quite closely in its implementation, since the heavy lifting, required for realizing efficiency boosts from matrix-free approaches, no longer interferes with the algorithm's implementation.
Most pitfalls, that are commonly associated with implementation optimization, are effectively tidied up from the abstraction provided by the \Matrix class-\gls{api}.
For an example of such an algorithm that greatly benefits from matrix-free transforms, see \Cref{usndt_example}.


\subsubsection{Architecture Remarks (under-the-hood section)}

It is well known that (despite its many sophisticated features) Python is not a language that is particularly known for rapid code execution.
To address this issue we make extensive use of Cython in order to reduce the overhead introduced by our object oriented model, following the principle of reducing Python interpreter and object model interaction where possible.
This is especially necessary, when matrix-free operators are to be implemented abstractly from combining many smaller (\Matrix-)operators of lower-complexity.
Quickly the situation is reached, where the high amount of embedded transform calls limits overall performance, if \gls{api} performance were left unconsidered.

To minimize call overhead, the complete class model architecture and all built-in class implementations utilize Cython to compile the code and statically-link it back into the Python \gls{api}.
This way the Python interpreter can be evaded not only during intra-package execution flow (e.g. method calls), but also for most interactions with Numpy (by means of its provided C-API), resulting in a significant performance boost over pure (interpreted) Python code.
Dodging the runtime penalty that comes with run-time dynamic typing, \fastmat internally makes heavy use of static typing and supports the data types integer (8, 16, 32 and 64 bit), float and complex (both for single and double precision).
Another runtime penalty arises from the creation of vast amounts of Views into `ndarray`s during internal array slicing operations.
This is mitigated by the use of a dedicated mutable striding operators on arrays, which greatly relieves stress on creating and garbage-collecting short-lived Python object.

As another aspect, building structured operators from (a large amount of) smaller operators may also result another computational disadvantage:
Instead of using the matrix-free algorithm to realize $\bm \phi$ or $\bm\beta$, it might be more efficient on modern CPUs to use the conventional matrix-vector ``dot'' product as a shortcut, especially when memory consumption is not a concern.
Since efficient transform outperform dot-products with increasing problem size, there usually exists a threshold problem size up until which the dense transform is more efficient.
The issue now is to determine easily where this threshold lies for a particular transform.

In \fastmat this process is called \emph{calibration} and is available for any \Matrix class (also user-defined ones) as part of the normal user interface.
For this, \fastmat offers simple-call means to one-time generate runtime estimation models (over problem size and the amount of vectors to process simultaneously) for $\bm\phi$ and $\bm\beta$.
During the instantiation of a \Matrix instance the runtime performance and that of a dense product are estimated if calibration data exists for the session.
When $\bm\phi$ or $\bm\beta$ is finally used, the optimal path for this particular set of vectors can immediately be chosen.
This way, \fastmat features integrated run-time computation path optimization.

%
%
%
%

All these considerations make \fastmat a very powerful abstraction toolbox for matrix-free structured linear operators.
However, the final challenge is to implement these features in a way that the following three goals are met together:
First, to maintain reusability in different contexts, a \Matrix instance must behave the same regardless of context as long it makes sense mathematically.
Therefore, in embracing the \emph{duck typing} mantra of Python, input and output sanitizing is required for all user-interface methods.
Secondly, it must be easily possible to define new or extend existing classes, regardless of them being built-in or user defined, or to override particular portions of any classes' take on the user interface.
And finally, readability, use- and reusability must be maintained for a user experience of lowest possible complexity.
Aside from the already discussed notation-code-duality, it must also be avoided that users repeatedly have to take care about not breaking the internal \gls{api} or functionality when overriding methods

These goals are achieved by adding an input preparation layer directly at the class-\gls{api} entry methods of the \Matrix baseclass.
Every class-\gls{api} method takes care of input compliance and applies more advanced features, such as pre-computation, result caching or runtime execution path optimization (\emph{calibration}).
The actual implementation code is provided in a separate private method.
Since the user only overrides these private methods when implementing custom structured matrices, interface consistency is always guaranteed, which greatly reduces code complexity for the user.

For properties, that apply pre-computation caching, the overall code execution (and implementation overloading) mechanism is as follows:
Assume one wishes to access the property \texttt{A.foo}.
There exists a shadow attribute \texttt{A.\_foo}, which caches the computation result, which is immediately returned if it is already present.
If not, the method \texttt{A.getFoo()} is triggered, which also serves as direct \gls{api} entry point for the computation of `foo`.
Identical to any other class-\gls{api} method, interface integrity is established by this entry method, which is why it must not be overloaded directly (to save the user having to tediously replicate its function every time).
Finally, \texttt{A.\_getFoo()} is called, which only executes the actual implementation.
Therefore, if a user wants to alter the behavior of \texttt{A.getFoo()}, overloading \texttt{A.\_getFoo()} with a new implementation is all that needs to be done.
This mechanism is used to cover all class-\gls{api} methods and properties, such as the storage of singular values or column norms.


%

%% file: figures/figure-baseclass.tex
\sisetup{
    table-figures-exponent = 2
}
\setlength{\tabcolsep}{2mm}
\renewcommand{\arraystretch}{1.3}
\begin{tabular}{@{} r c | T T @{}}
    Feature
        & Notation
        & Operator
        & \Matrix Attribute
        \\
    \hline
    Dense representation
        & $\bm M$
        & \texttt{M[...]}
        & \texttt{A.array}
        \\
    Element Indexing$^\ast$
        & $\bm M_{i, j}$
        & \texttt{M[i, j]}
        & \texttt{A.getItem(i, j)}
        \\
    Row Access$^\ast$
        & $\bm M_{i, \bullet}$
        & \texttt{A[row, :]}
        & \texttt{A.getRow / A.getRows(i)}
        \\
    Column Access$^\ast$
        & $\bm M_{\bullet, j}$
        & \texttt{A[:, col]}
        & \texttt{A.getCol / A.getCols(j)}
        \\
    Row Norms
        & $\frac{\bm M_{i, \bullet}}{||M_{i, \bullet}||}$
        &
        & \texttt{A.rowNorms}
        \\
    Column Norms
        & $||M_{\bullet, j}||$
        &
        & \texttt{A.colNorms}
        \\
    Row Normalization
        & $||M_{i, \bullet}||$
        &
        & \texttt{A.rowNormalized}
        \\
    Column Normalization
        & $\frac{\bm M_{\bullet, j}}{||M_{\bullet, j}||}$
        &
        & \texttt{A.colNormalized}
        \\
    Transpose
        & $\bm M^T$
        &
        & \texttt{M.T}
        \\
    Hermitian Transpose
        & $\bm M^H$
        &
        & \texttt{M.H}
        \\
    Complex Conjugate
        & $\text{conj}(\bm M)$
        &
        & \texttt{M.conj}
        \\
    Inverse
        & $\bm M^{-1}$
        &
        & \texttt{M.inverse}
        \\
    Pseudoinverse
        & $\bm M^+$
        &
        & \texttt{A.pseudoInverse}
        \\
    Gramian
        & $\bm M^H \bm M$
        &
        & \texttt{A.gram}
        \\
    Scipy Linear Operator
        &
        &
        & \texttt{A.scipyLinearOperator}
        \\
    \bottomrule
    \multicolumn{4}{l}{\scriptsize $^\ast$\textit{Does not offer internal caching by the \texttt{\_}-prefix syntax.}}
\end{tabular}

%% file: figures/figure-matrices.tex
\newcommand\matrixImage[1]{%
    \includegraphics[height=32mm]{figures/#1.pdf}
}
\begin{subfigure}[t]{0.99\linewidth}
    \begin{minipage}[t]{0.24\linewidth}
        \centering
        \matrixImage{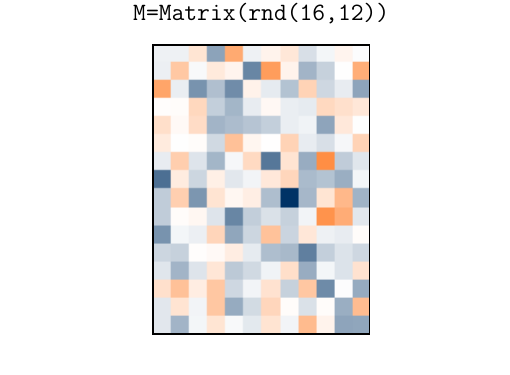}
        \\
        \matrixImage{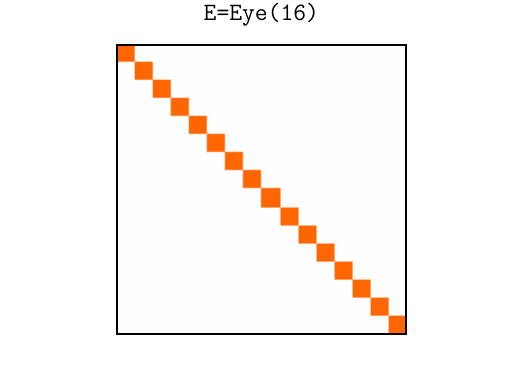}
        \\
        \matrixImage{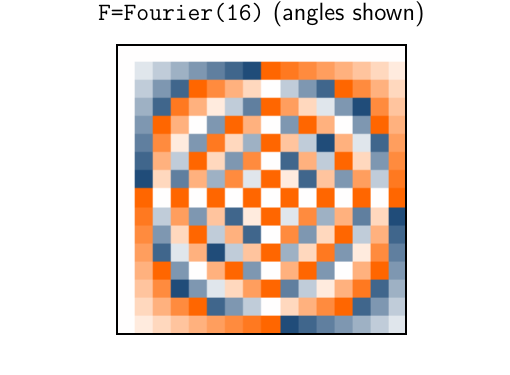}
        \\
        \matrixImage{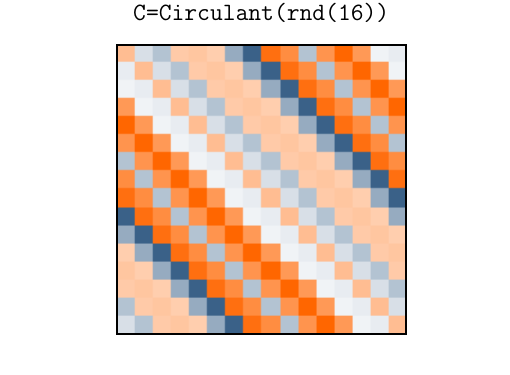}
    \end{minipage}
    \hfill
    \begin{minipage}[t]{0.24\linewidth}
        \centering
        \matrixImage{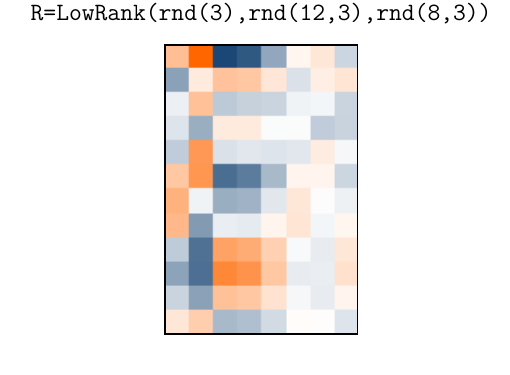}
        \\
        \matrixImage{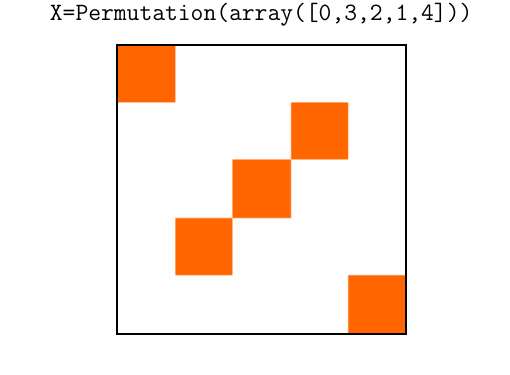}
        \\
        \matrixImage{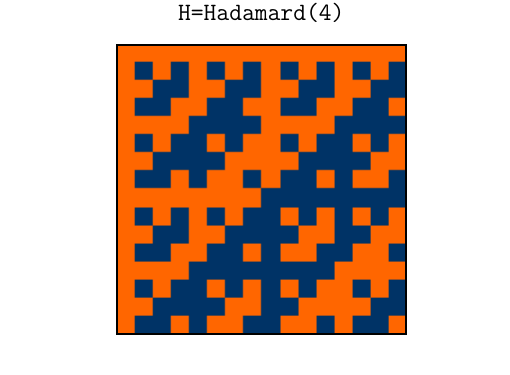}
        \\
        \matrixImage{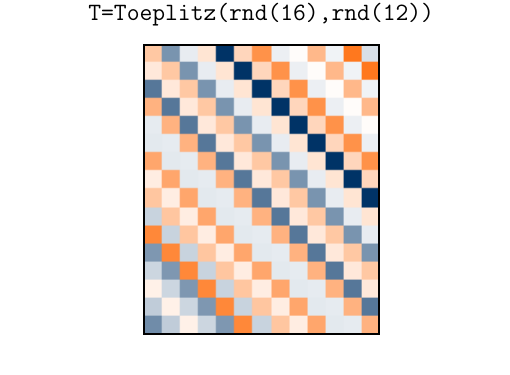}
    \end{minipage}
    \hfill
    \begin{minipage}[t]{0.24\linewidth}
        \centering
        \matrixImage{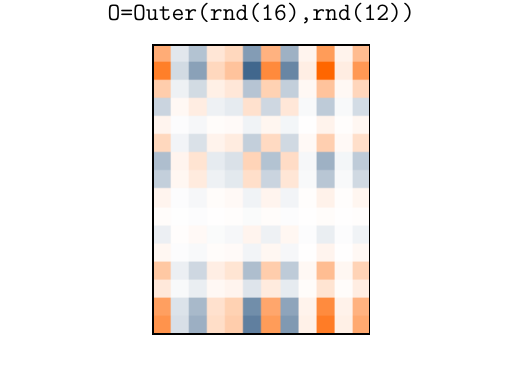}
        \\
        \matrixImage{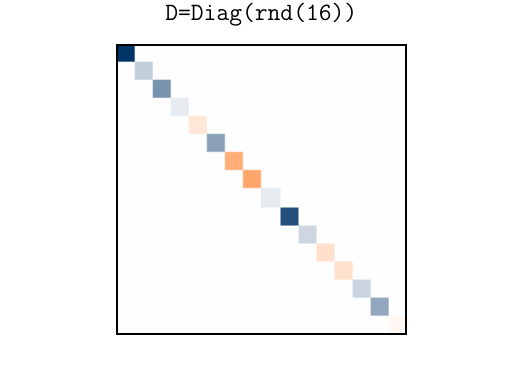}
        \\
        \matrixImage{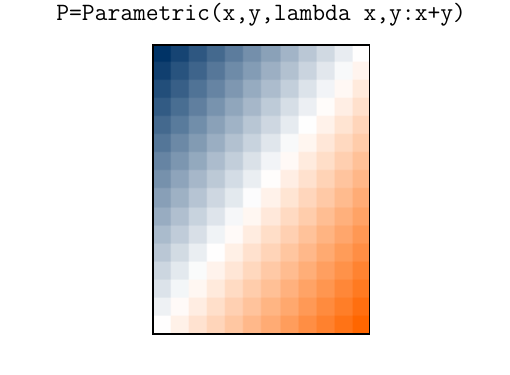}
        \\
        \matrixImage{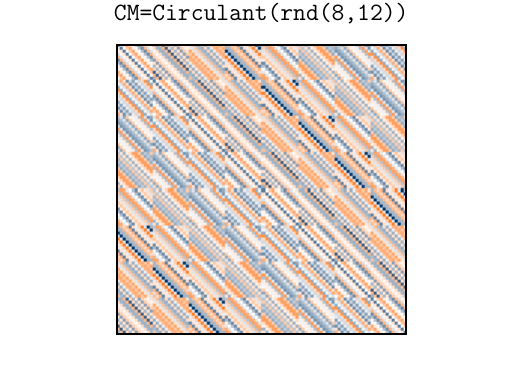}
    \end{minipage}
    \hfill
    \begin{minipage}[t]{0.24\linewidth}
        \centering
        \matrixImage{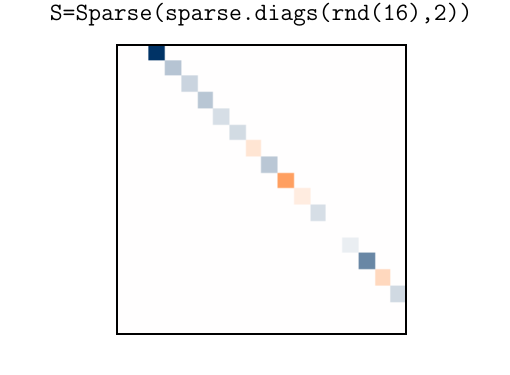}
        \\
        \matrixImage{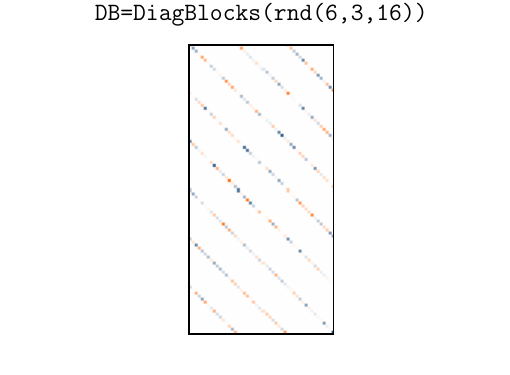}
        \\
        \matrixImage{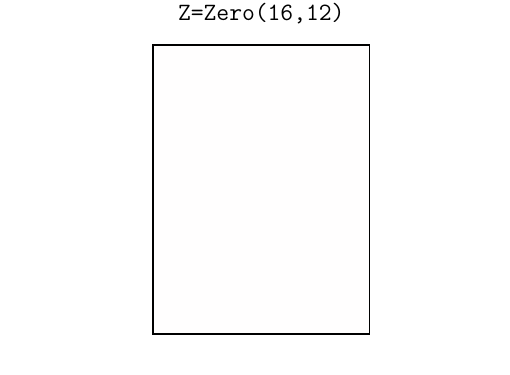}
        \\
        \matrixImage{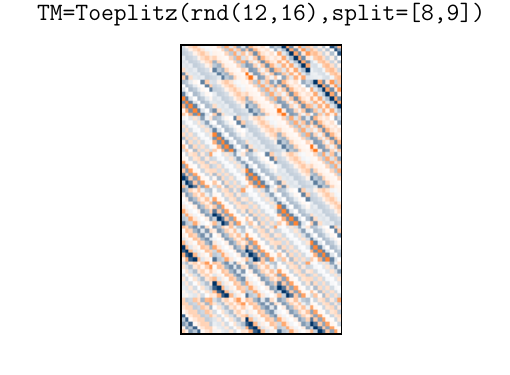}
    \end{minipage}
\end{subfigure}%

\begin{subfigure}[t]{0.99\linewidth}
    \begin{minipage}[t]{0.24\linewidth}
        \centering
        \matrixImage{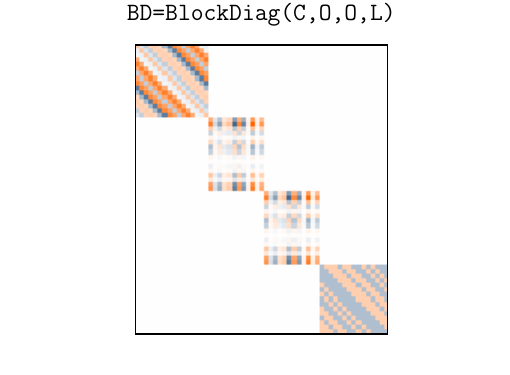}
        \\
        \matrixImage{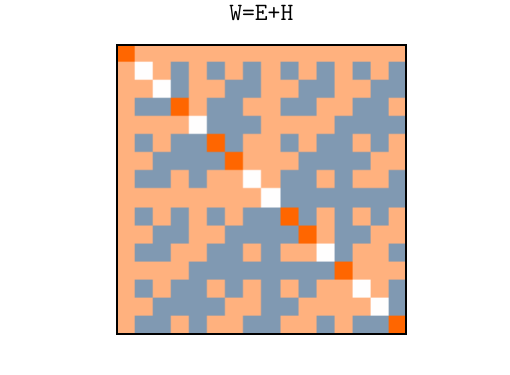}
    \end{minipage}
    \hfill
    \begin{minipage}[t]{0.24\linewidth}
        \centering
        \matrixImage{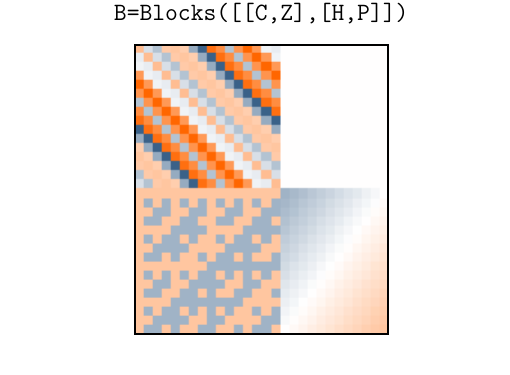}
        \\
        \matrixImage{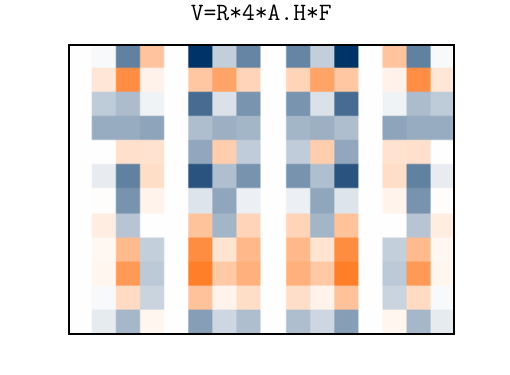}
    \end{minipage}
    \hfill
    \begin{minipage}[t]{0.24\linewidth}
        \centering
        \matrixImage{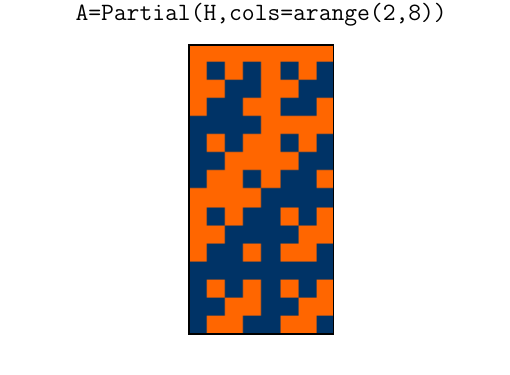}
        \\
        \matrixImage{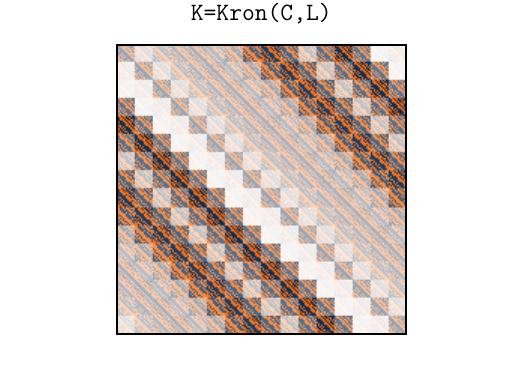}
    \end{minipage}
    \hfill
    \begin{minipage}[t]{0.24\linewidth}
        \centering
        \matrixImage{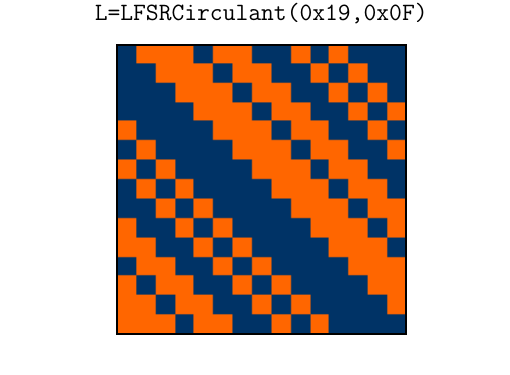}
    \end{minipage}
\end{subfigure}
\vspace{-1cm}

%% file: sections/examples.tex

%

Here, we first give a short overview on how to define a custom \Matrix implementation and second, we demonstrate an example for sparsity driven signal processing in nondestructive testing using ultrasound.

\subsection{Defining your Custom Matrix-free Operators and Algorithms}

\begin{figure}[t]
  \centering
  \input{figures/figure-harold}
  \caption{Our conceptual engineer Harold is relieving some pain at work}
  \label{harold}
\end{figure}

Based on what we have explained in \Cref{general_baseclass}, we will now see how these concepts can be put to work for Harold.
Having suffered the pain of optimizing structured numeric implementations in the past, Harold is eager to try the concept for the implementation of a $\bm 1$ matrix of shape $N \times M$, which is defined such that every one of its elements equals to $1$.
In the following, we will show step-by-step how to arrive at an efficient matrix-free implementation using \fastmat (see~\Cref{harold}).

At first, Harold has to identify an already implemented class, that is already close to his desired outcome.
Often, inheriting from the right baseline already saves most of the work.
For our example, the \Matrix baseclass is already best suited.
This cherry-picking can also be combined by a suitable composition of multiple \Matrix classes.
After this decision, Harold starts with incrementally specializing his implementation by first overriding the methods \texttt{\_forward($\bullet$)} and \texttt{\_backward($\bullet$)} (corresponding to $\bm \phi$ and $\bm \beta$) to define the general behavior.
Also, the \texttt{\_\_init\_\_($\bullet$)} method must be specified and it must either end in a call to \texttt{initProperties($\bullet$)} for initializing the class-\gls{api} from scratch, or in a \texttt{super($\bullet$)} reference to the parent's class \texttt{\_\_init\_\_($\bullet$)} method.
If certain pre-computations should be carried out, they should also be resided in the body of the \texttt{\_\_init\_\_($\bullet$)} method.
Later on, Harold can step-by-step override other required properties, like the efficient calculation of singular values or eigenvectors, if he knows ways to improve these over the by-default inherited general implementations.

\begin{listing}[hbt!]
  \begin{singlespace}
    \begin{minted}[fontsize=\small,linenos,frame=leftline]{py}
  import numpy, fastmat

  class Ones(fastmat.Matrix):
      '''A simple 1-Matrix implementation'''

      def __init__(self, numRows, numCols):
          '''Initialize fastmat class-API and define shapes'''
          self._initProperties(numRows, numCols, numpy.int8)

      def _forward(self, arrX):
          '''Define forward transform'''
          return numpy.resize(
              numpy.sum(arrX, axis=0, keepdims=True),
              (self.numRows, arrX.shape[1])
          )

      def _backward(self, arrX):
          '''Define backward transform'''
          return numpy.resize(
              numpy.sum(arrX, axis=0, keepdims=True),
              (self.numCols, arrX.shape[1])
          )

      ### Additional methods required to be defined
      ###  ... for Testing:     reference()      and _getTest()
      ###  ... for Calibration: _getComplexity() and _getBenchmark()

  # ############################# Instantiate and Use the new class
  N, M = 4, 5                ; O = Ones(N, M)
  xf = numpy.random.randn(M) ; xb = numpy.random.randn(N)
  print(xf, '\n', O * xf)    ; print(xb, '\n', O.H * xb)
    \end{minted}
  \end{singlespace}
  \caption{Example of custom matrix-free operator implementation and test case definition}
  \label{classexample}
\end{listing}

See \Cref{classexample} for a working minimal example.
Please note that, due to the limited size of this document, the setup of the test- and calibration subsystems is not included in this example, as these require the specification of additional methods to function.
Testing requires the specification of a dense reference array (to compare against) with the \texttt{reference()} method and the description of a test case generator from \texttt{\_getTest()}.
For calibration to work, a model for estimation of the relative complexity of $\bm \phi$ and $\bm \beta$ for a particular instance must be returned from \texttt{\_getComplexity()}.
During the actual calibration process, which is invoked by calling \texttt{fastmat.core.calibrateClass($\bullet$)}, a Benchmark probe on $\bm \phi$, $\bm \beta$ and the call overhead for increasing problem sizes is performed.
This requires the definition of a benchmark case generator in \texttt{\_getBenchmark()}, which defines at least the \texttt{forward} and \texttt{overhead} benchmark types.
Then, \fastmat is able to automatically calibrate this class and use this information to optimize the execution path during runtime.
The Benchmark results are then fitted to the provided calibration model to extract the model parameters.
The resulting calibration model is available for the session and may also be saved to or loaded from the file system.

Please consult the public package documentation on more detailed information on how to setup and use these subsystems.

\subsection{Application to Ultrasonic Nondestructive Testing}\label{usndt_example}

Ultrasound is used in a wide range of fields such as manufacturing, medicine or maintenance as a nondestructive modality to characterize the interior of an object under test.
In non-destructive testing one is further interested in detecting and characterizing possible defects within a specimen without afflicting any damage or alterations to the object.
Following, we summarize an example of recent research carried out on the 3D reconstruction of defects using ultrasonic non-destructive testing~\cite{journal-tuffc-2021} with special focus on its implementation using matrix-free methods and the proposed package \texttt{fastmat}.

A transducer, emitting ultrasonic pulses into the specimen, is moved along a predefined trajectory to collect the measurement data.
At each position, we take snapshots by inserting a known waveform and recording the echoes originating at material boundaries, due to the change in acoustic impedance.
This process, as depicted in \Cref{measurement_setup}, creates a synthetic aperture, which allows us to recover the (previously unknown) locations of such rapid impedance changes in the investigated medium.
These usually indicate interesting areas, which we aim to extract from the measurements.

\begin{figure}[t]
  \input{tikz/measurement_setup}
  \caption{Transducer path and the measurement positions.}
  \label{measurement_setup}
\end{figure}

In order to formulate a tractable parameter estimation problem, we first introduce a linear model, which maps a vector encoding possible defect locations onto a possible observation captured by the transducer.
Very generally, this reads as
\begin{equation}\label{measurement_model}
	\bm b = \bm H \cdot \bm x + \bm n,
\end{equation}
where $\bm b, \bm x \in \mathbb{C}^n$ and $\bm H \in \mathbb{C}^{n \times n}$.
In this equation, $\bm b$ is the obtained measurement from the transducer as in \Cref{measurement_setup}.

The matrix $\bm H$ models our assumptions of the physical process taking place in the medium: if we insert pulsed waveforms, these pulses get reflected by possible defects and can be seen in the recorded echos.
Based on this model, the entries in $\bm x$ correspond to certain positions in the object. A nonzero entry indicates a reflector or defect at said position. Consequently, we wish to recover $\bm x$ from the measurements $\bm b$ in order to infer these positions.

The transducer moves over points that are aligned on a regular $2D$ grid located in a hyperplane of $\mathbb{R}^3$.
In case when we introduce a well chosen grid for the possible defect locations, one can show that the matrix
\[
 \bm H = \left[\bm H_{i,j}\right]_{i,j = 1}^{N_1},
\]
is a block matrix, where each block $\bm H_{i,j} \in \mathbb{C}^{(2N_2 - 1)(2 N_3 - 1) \times (2N_2 - 1)(2 N_3 - 1)}$ is a $2$-level Toeplitz matrix.
Since $N_1$, $N_2$ and $N_3$ scale with the spatial resolution during measurement and reconstruction of the defects locations\rev{, as they constitute the number of samples in each of the spatial directions. T}he dense matrix $\bm H$ quickly assumes an insane amount of elements ($N_1^2(2N_2 - 1)^2(2 N_3 - 1)^2$).
Since this is simply too much for current system memory and computation resources we have to employ a matrix-free implementation of the transforms involving the dense matrix $\bm H$ to stay within the realms of practical feasibility.

Now, consider a single $\bm H_{i,j}$, which due to its structure represents a $2$D non-cyclic convolution.
Such a convolution can be implemented by means of a larger cyclic convolution and appropriate zero-padding of the input.
Leaving out technical details, one can show that
\begin{equation}\label{H_ij}
	\bm H_{i,j} = \mathcal{P}\left(
		(\bm F_{2N_2 - 1} \otimes \bm F_{2N_3 - 1})
		\cdot \Diag(\bm h_{i,j})
		\cdot (\bm F_{2N_3 - 1}^H \otimes \bm F_{2N_2 - 1}^H)
	\right),
\end{equation}
where $\otimes$ denotes the matrix Kronecker product, $\Diag(\bm x)$ is a diagonal matrix with $\bm x$ as its diagonal and $\mathcal{P}$ is the operator in charge of the correct zero-padding.
If we want to implement this matrix in the herein described package one would use a suitable combination of \texttt{Partial}, \texttt{Kron}, \texttt{Diag}, \texttt{Product}, \texttt{Hermitian} and \texttt{Fourier}.

In order to carry out the estimation of $\bm x$ based on $\bm H$ and $\bm b$, one popular approach is the following optimization problem:
\begin{equation}\label{ell1_min}
	\Min\limits_{\bm x}
		\Vert \bm H \bm x - \bm b \Vert_2^2
		+ \tau \Vert \bm x \Vert_1,
\end{equation}
where $\Vert \bm x \Vert_p = (\sum_i \vert x_i \vert^p)^{(1/p)}$ and $\tau > 0$ being some suitably chosen hyper parameter.
Interestingly, there are algorithms to approximately solve problems like in \eqref{ell1_min} efficiently, solely based on matrix-vector products.
It has been discovered in recent years, that the regularizing term $\tau \Vert \bm x \Vert_1$ is able to enforce \emph{sparsity} in $\bm x$, which corresponds to the assumption that there are only a few defects within the specimen under test.

One approach to solve the problem in \eqref{ell1_min} is the \gls{ista}\rev{~\cite{beck2010FastIterativeShrinkage}}, which is the basic inspirations for a whole array of similar algorithms with varying performance. However, for the sake of simplicity, we present the simple and basic version.
For a given initial estimate $\bm x_0 \in \mathbb{C}^n$ we simply iterate the following expressions until we have satisfied some stopping criterion:
\[
	\bm x_{k+1}
		= \bm\tau_{s}\left(
			\bm H^H \cdot (\bm H \cdot \bm x_k - \bm b)
	\right),
\]
where $\bm \tau_s : \mathbb{C}^n \rightarrow \mathbb{C}^n$ is a non-linear function (called the soft thresholding operator) to the level $s > 0$, which exhibits linear runtime over the size of $\bm x$.
This means the largest computational effort for \gls{ista} resides in the computation of the $\bm\phi_{\bm H}$ and $\bm\beta_{\bm H}$ projections respectively, rendering it a perfect application for a matrix-free implementation.
For \gls{ista} we simply end up with
\[
	\bm x_{k+1}
		= \bm\tau_{s}\left(
			\bm\beta_{\bm H}(\bm\phi_{\bm H}(\bm x_k) - \bm b)
	\right).
\]
%
\begin{figure}[t]
  \input{tikz/muse_sketch.tex}
  \input{tikz/FISTA_reco_MUSE_axis0_largerScenario.tex}
  \input{tikz/OMP_reco_MUSE_axis0_largerScenario.tex}
  \input{tikz/SAFT_reco_MUSE_axis0_largerScenario.tex}
  \caption{\emph{Sparse recovery results for a steel specimen (top view)} -- Top to bottom: Sketch of the specimen (Top View), \gls{fista} reconstruction, \gls{omp} reconstruction, beamforming via \gls{saft}.
  \label{MUSElarge}
}
\end{figure}
In research on ultrasonic non-destructive testing, a common procedure is to compare imaging algorithms using a test specimen with artificially introduced target defects.
Such defects are commonly represented by flat holes that are drilled into the bottom of the specimen.
In \Cref{MUSElarge} we show some examples for possible reconstructions of such defects in a steel block.
In this case we use the transducer to localize these defects by measuring into the specimen from the top side.
See \Cref{MUSElarge} for results of three different algorithms, namely \gls{fista}\rev{~\cite{beck2010FastIterativeShrinkage}}, a slight variant of the described \gls{ista}, \gls{omp}\rev{~\cite{tropp2007signal}} (another sparsity-enforcing algorithm well suited for matrix-free methods), and  \gls{saft}~\cite{spies_jnde_2012}, a beamforming method that simply computes $\hat{\bm x} = \bm \beta_{\rev{\bm H}}(\bm b)$.

The fact, that the modeling stage and the algorithm design stage remain completely separate from each other, renders the proposed toolbox very advantageous as a research tool.
The same matrix-free representation can easily be plugged into other suitable algorithms, which only have to make use of the interface described before.

\begin{figure*}
    \input{tikz/muse_global_random_dm_df_single_coefficient.tex}
    \caption{\rev{Top and side view of a FISTA reconstruction using $20$ steps, $n_f=1$ Fourier coefficients and $\bm \Psi \bm H$ as the system matrix.}}
    \label{muse_side}
\end{figure*}

On the other hand, we can also easily adapt for a different measurement scenario.
\gls{cs} uses the techniques developed in \gls{ssr} for estimating signal parameters from compressed measurements.
If we apply this approach to the problem of defect detection we get a different observation model, which reads as
\begin{equation}\label{compressed_measurement_model}
	\hat{\bm b} = \bm \Psi \cdot \bm H \cdot \bm x + \bm n,
\end{equation}
where the matrix $\bm \Psi \in \mathbb{C}^{m \times n}$ is such that $m \ll n$, resulting in a $\hat{\bm b}$, containing far fewer measurements than $\bm b$.

In our case we use $\bm \Psi$ to compress the $N_2 \cdot N_3$ many single-pulse-echo-measurements independently in frequency domain.
This way we get
\[
\bm \Psi = \BlkDiag\{\bm \Phi_{1,1}, \dots, \bm \Phi_{N_1,N_2}\},
\]
where each $\bm \Phi_{i,j}$ is a matrix containing a random selection of $n_f \in \mathbb{N}$ rows of the Fourier matrix and the $\BlkDiag$ operator arranges the matrices in a block-diagonal structure.
Clearly, the measurement matrix $\bm \Psi$ has a matrix-free representation and as such also does the product $\bm \Psi \bm H$.
If we simply plug the new measurements $\hat{\bm b}$ and the product $\bm \Psi \bm H$ into the already discussed algorithms like \gls{omp}, \gls{fista} or \gls{ista}, we get results as depicted in \Cref{muse_side}.
There, we only used a single Fourier coefficient from each of the originally collected pulse-echo-measurements.

As proof of our claim about the resulting narrow relationship of code and mathematical notation when using \texttt{fastmat}, we show a simplified example implementation of \rev{\eqref{H_ij} and \eqref{ell1_min}} in \Cref{codeexample}.

\begin{listing}[bt]
  \begin{singlespace}
    \begin{minted}[fontsize=\small,linenos,frame=leftline]{py}
  # either you could go by
  # coding Eq. (7) literally,
  FK = fastmat.Kron(fastmat.Fourier(N2), fastmat.Fourier(N3))
  H = fastmat.Blocks([
    [ fastmat.Partial(
      FK * fastmat.Diag(h[i, j]) * FK.H, cols=vec_zeropad
    )
      for j in range(N1)
    ] for i in range(N1)
  ])

  # or you could just use the
  # built-in Multi-Level Toeplitz class
  H = fastmat.Blocks([
    [ fastmat.Toeplitz(h[i, j])
      for j in range(N1)
    ] for i in range(N1)
  ])

  # Using H in an algorithm (8) works as follows:
  import fastmat.algorithms
  alg = fastmat.algorithms.ISTA(H, numLambda=lambda)
  x = alg.process(b)
    \end{minted}
  \end{singlespace}
  \caption{The code implementation for \Cref{H_ij} (two variants given) and \Cref{ell1_min}}
  \label{codeexample}
\end{listing}

This second example underlines how the proposed toolbox not only separates the modeling and the algorithm design stage, but also to allow further abstraction of the model shown by easily adding a compressive data acquisition model to the design.
A suitable model and algorithm can be tested with different (simulated) sampling schemes, with the structure of the measurement setup also being reflected in the implementation.

%% file: figures/figure-harold.tex
\begin{subfigure}[b]{0.49\linewidth}
    \centering
    \includegraphics[width=0.95\linewidth]{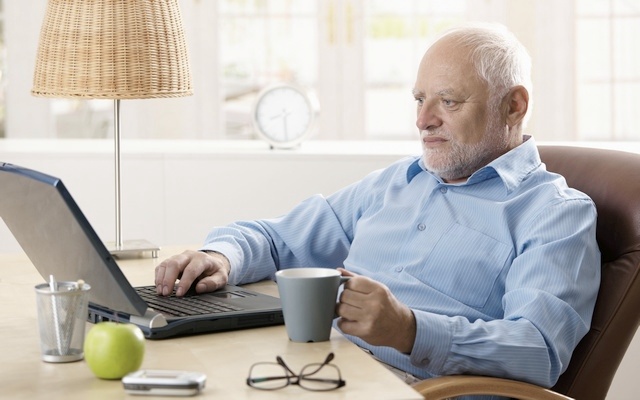}
    \caption{Lost in layers of structural optimization}
\end{subfigure}
\hfill
\begin{subfigure}[b]{0.49\linewidth}
    \centering
    \includegraphics[width=0.95\linewidth]{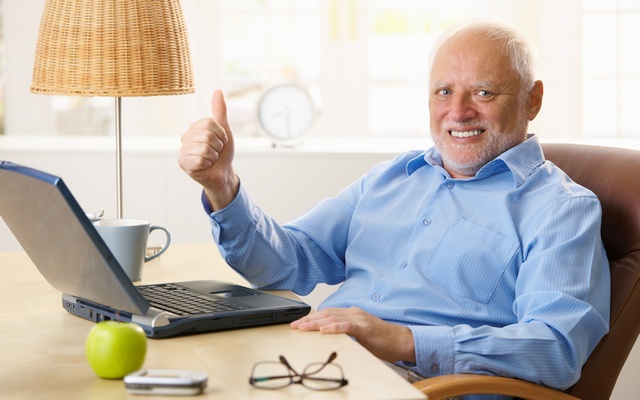}
    \caption{Harold discovered \texttt{pip install fastmat}}
\end{subfigure}

%% file: tikz/measurement_setup.tex

\newcommand{\Nx}{3}
\newcommand{\Ny}{2}
\newcommand{\dx}{2cm}
\newcommand{\dy}{2cm}
\newcommand{\dz}{2cm}
\newcommand{\rCircle}{0.055cm}

\newcommand{\drawTransducer}[4]{ 
	\draw[draw = none, fill = white] (axis cs: #3 - #1, #4 - #2, 0) -- (axis cs: #3 + #1, #4 - #2, 0) -- (axis cs: #3 + #1, #4 + #2, 0) --  (axis cs: #3 + #1, #4 + #2, -#1) -- (axis cs: #3 - #1, #4 + #2, -#1) -- (axis cs: #3 - #1, #4 - #2, -#1) -- (axis cs: #3 - #1, #4 - #2, 0);
	\draw[] (axis cs: #3 - #1, #4 - #2, 0) -- (axis cs: #3 + #1, #4 - #2, 0) -- (axis cs: #3 + #1, #4 - #2, -#1) -- (axis cs: #3 - #1, #4 - #2, -#1) -- (axis cs: #3 - #1, #4 - #2, 0);
	\draw[] (axis cs: #3 + #1, #4 - #2, 0) -- (axis cs: #3 + #1, #4 + #2, 0) -- (axis cs: #3 + #1, #4 +  #2, -#1) -- (axis cs: #3 - #1, #4 +  #2, -#1) -- (axis cs: #3 - #1, #4 - #2, -#1);
	\draw[] (axis cs: #3 + #1, #4 - #2, -#1) -- (axis cs: #3 + #1, #4 +  #2, -#1);
}

	\begin{tikzpicture}[]
		\begin{axis}[height=20cm, width=40cm,
					 ticks=none,axis lines=center,
             		 xlabel={\footnotesize $x$},
             		 ylabel={\footnotesize $y$},
             		 zlabel={\footnotesize $z$},
					 x={(\dx,0cm)},
					 y={(0.25cm,0.30cm)},
					 z={(0cm,-\dz)},
					 xmin = -0.12,
					 xmax = 6,
					 ymin = -0.3,
					 ymax = 4,
					 zmin = -0.5,
					 zmax = 1.2,
					 z label style={at={(axis cs: 0,0,1.2)},anchor=east},
            		 ]

        	\foreach \yValue in {1,...,\Ny} {
        		\foreach \xValue in {1,...,\Nx} {
    				\edef\temp{\noexpand\draw[color = black, fill = black, anchor =center] (axis cs: \xValue + 1, \yValue + 1, 0) circle(\rCircle);}
    				\temp
    			}
			}
			\foreach \yValue in {1,...,\Ny} {
        		\foreach \xValue in {0} {
    				\edef\temp{\noexpand\draw[color = black, fill = black, anchor =center] (axis cs: \xValue + 1, \yValue + 1, 0) circle(\rCircle);}
    				\temp
    			}
			}
			\foreach \yValue in {0} {
        		\foreach \xValue in {1,...,\Nx} {
    				\edef\temp{\noexpand\draw[color = black, fill = black, anchor =center] (axis cs: \xValue + 1, \yValue + 1, 0) circle(\rCircle);}
    				\temp
    			}
			}

			\foreach \yValue in {0,2} {
        		\foreach \xValue in {1,...,3} {
    				\edef\temp{\noexpand\draw[color = black, dotted, ->, shorten >=1mm, shorten <=1mm,] (axis cs: \xValue + 0, \yValue + 1, 0) -- (axis cs: \xValue + 1, \yValue + 1, 0);}
    				\temp
    			}
			}

			\foreach \yValue in {1} {
        		\foreach \xValue in {1,...,3} {
    				\edef\temp{\noexpand\draw[color = black, dotted, ->, shorten >=1mm, shorten <=1mm,] (axis cs: \xValue + 1, \yValue + 1, 0) -- (axis cs: \xValue + 0, \yValue + 1, 0);}
    				\temp
    			}
			}

			\draw[color = black, dotted, ->, shorten >=1mm, shorten <=1mm,] (axis cs: 3 + 1, 0 + 1, 0) -- (axis cs: 3 + 1, 1 + 1, 0);

        	\draw[color = black, dotted, ->, shorten >=1mm, shorten <=1mm,] (axis cs: 0 + 1, 1 + 1, 0) -- (axis cs: 0 + 1, 2 + 1, 0);

			\drawTransducer{0.5}{0.25}{1}{1}
			\draw[snake=expanding waves]   (axis cs: 1,1,0.2) -- (axis cs: 1,1,1.5);

			\node[coordinate] (x_annot_start_real) at (axis cs: 2,1,0) {};
			\node[coordinate] (x_annot_stop_real) at (axis cs: 3,1,0) {};

			\node[coordinate, below = 0.3cm of x_annot_start_real] (x_annot_start) {};
			\node[coordinate, below = 0.3cm of x_annot_stop_real] (x_annot_stop) {};

			\draw[color = black,  <->] (x_annot_start) -- node[below] {\footnotesize$\Delta x$} (x_annot_stop);

			\node[coordinate] (y_annot_start_real) at (axis cs: 4,0.8,0) {};
			\node[coordinate] (y_annot_stop_real) at (axis cs: 4,1.8,0) {};

			\node[coordinate, right = 0.4cm of y_annot_start_real] (y_annot_start) {};
			\node[coordinate, right = 0.4cm of y_annot_stop_real] (y_annot_stop) {};

			\draw[color = black,  <->] (y_annot_start) -- node[right] {\footnotesize\hspace{0.1cm}$\Delta y$} (y_annot_stop);

		\end{axis}
	\end{tikzpicture}

%% file: tikz/muse_sketch.tex
\newcommand{\offsetx}{0}
\newcommand{\scalingx}{20.8}
\newcommand{\offsety}{80}
\newcommand{\scalingy}{-20}
\newcommand{\circlesTwomm}{2}
\newcommand{\circlesThreemm}{3}
\newcommand{\circlesFivemm}{5}
\begin{tikzpicture}
		\begin{axis}
		[
	enlargelimits = false,
    axis on top = true,
    axis equal image,
		width=0.9\linewidth,
    ylabel = {\small mm},
    xtick = {0, 180, 361},
    xticklabels = {},
    ytick = {0,25,50},
    yticklabels = {0, 12.5, 25},
		]
		 \addplot graphics [
        xmin = 0.000000,
        xmax = 361.000000,
        ymin = 0.000000,
        ymax = 70.000000
    ]{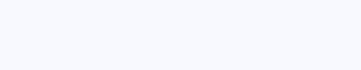};
		\draw (axis cs: \offsetx + \scalingx * 13.84,\offsety + \scalingy * 2.735) circle (\circlesTwomm);
		\draw (axis cs: \offsetx + \scalingx * 14.34,\offsety + \scalingy * 2.235) circle (\circlesTwomm);
		\draw (axis cs: \offsetx + \scalingx * 14.84,\offsety + \scalingy * 1.735) circle (\circlesTwomm);

		\draw (axis cs: \offsetx + \scalingx * 12.84,\offsety + \scalingy * 2.735) circle (\circlesThreemm);
		\draw (axis cs: \offsetx + \scalingx * 13.34,\offsety + \scalingy * 2.235) circle (\circlesThreemm);
		\draw (axis cs: \offsetx + \scalingx * 13.84,\offsety + \scalingy * 1.735) circle (\circlesThreemm);

		\draw (axis cs: \offsetx + \scalingx * 11.34,\offsety + \scalingy * 2.735) circle (\circlesFivemm);
		\draw (axis cs: \offsetx + \scalingx * 12.34,\offsety + \scalingy * 1.735) circle (\circlesFivemm);

		\draw[rounded corners=2.0,rotate around={-45:(axis cs: \offsetx + \scalingx * 8.79,\offsety + \scalingy * 2.735)}] (axis cs: \offsetx + \scalingx * 8.64,\offsety + \scalingy * 3.225) rectangle (axis cs: \offsetx + \scalingx * 8.94,\offsety + \scalingy * 2.225) {};
		\draw[rounded corners=2.0,rotate around={-45:(axis cs: \offsetx + \scalingx * 9.79,\offsety + \scalingy * 1.735)}] (axis cs: \offsetx + \scalingx * 9.64,\offsety + \scalingy * 2.225) rectangle (axis cs: \offsetx + \scalingx * 9.94,\offsety + \scalingy * 1.225) {};

		\draw[rounded corners=2.0] (axis cs: \offsetx + \scalingx * 5.69,\offsety + \scalingy * 3.215) rectangle (axis cs: \offsetx + \scalingx * 5.99,\offsety + \scalingy * 1.215) {};

		\draw[rounded corners=2.0] (axis cs: \offsetx + \scalingx * 0.69,\offsety + \scalingy * 2.375) rectangle (axis cs: \offsetx + \scalingx * 2.69,\offsety + \scalingy * 2.075) {};

		\node[anchor=west] at (axis cs: \offsetx + \scalingx * 13.15,\offsety + \scalingy * 3.2)  {\footnotesize $\varnothing\SI{2}{\milli\metre}$};
		\node[anchor=west] at (axis cs: \offsetx + \scalingx * 13.04,\offsety + \scalingy * 0.87)  {\footnotesize $\varnothing\SI{3}{\milli\metre}$};
		\node[anchor=west] at (axis cs: \offsetx + \scalingx * 10.0,\offsety + \scalingy * 3.27)  {\footnotesize $\varnothing\SI{5}{\milli\metre}$};

		\node[anchor=west] at (axis cs: \offsetx + \scalingx * 5.4,\offsety + \scalingy * 3.5)  {\footnotesize $\leftrightarrow\SI{3}{\milli\metre}$};
		\node[anchor=west] at (axis cs: \offsetx + \scalingx * 2.75,\offsety + \scalingy * 2.255)  {\footnotesize $\updownarrow\SI{3}{\milli\metre}$};

		\end{axis}
\end{tikzpicture}

%% file: tikz/FISTA_reco_MUSE_axis0_largerScenario.tex
\begin{tikzpicture}
\begin{axis}[
    enlargelimits = false,
    axis on top = true,
    axis equal image,
    width=0.9\linewidth,
    ylabel = {\small mm},
    xtick = {0, 180, 361},
    xticklabels = {},
    ytick = {0,25,50},
    yticklabels = {0, 12.5, 25},
    colormap={FRI_white_blue_red}{
        rgb(0.0 pt)=(0.9686,0.9765,0.9922),
        rgb(0.010101010101010102 pt)=(0.9573435294117647,0.9696105882352941,0.9886925490196078),
        rgb(0.020202020202020204 pt)=(0.9404588235294118,0.9592764705882353,0.9834313725490196),
        rgb(0.030303030303030304 pt)=(0.9292023529411765,0.9523870588235295,0.9799239215686274),
        rgb(0.04040404040404041 pt)=(0.9123176470588236,0.9420529411764706,0.9746627450980392),
        rgb(0.050505050505050504 pt)=(0.9010611764705883,0.9351635294117647,0.971155294117647),
        rgb(0.06060606060606061 pt)=(0.8841764705882353,0.9248294117647059,0.9658941176470588),
        rgb(0.0707070707070707 pt)=(0.8672917647058824,0.9144952941176471,0.9606329411764706),
        rgb(0.08080808080808081 pt)=(0.8560352941176471,0.9076058823529412,0.9571254901960784),
        rgb(0.09090909090909091 pt)=(0.8391505882352941,0.8972717647058823,0.9518643137254902),
        rgb(0.10101010101010101 pt)=(0.8278941176470589,0.8903823529411765,0.948356862745098),
        rgb(0.1111111111111111 pt)=(0.8110094117647059,0.8800482352941177,0.9430956862745098),
        rgb(0.12121212121212122 pt)=(0.794124705882353,0.8697141176470589,0.9378345098039216),
        rgb(0.13131313131313133 pt)=(0.7828682352941176,0.8628247058823529,0.9343270588235294),
        rgb(0.1414141414141414 pt)=(0.7659835294117647,0.8524905882352941,0.9290658823529412),
        rgb(0.15151515151515152 pt)=(0.7547270588235294,0.8456011764705883,0.925558431372549),
        rgb(0.16161616161616163 pt)=(0.7378423529411765,0.8352670588235295,0.9202972549019608),
        rgb(0.1717171717171717 pt)=(0.7265858823529412,0.8283776470588236,0.9167898039215686),
        rgb(0.18181818181818182 pt)=(0.7097011764705883,0.8180435294117647,0.9115286274509804),
        rgb(0.1919191919191919 pt)=(0.6928164705882354,0.8077094117647059,0.9062674509803921),
        rgb(0.20202020202020202 pt)=(0.6815599999999999,0.80082,0.90276),
        rgb(0.21212121212121213 pt)=(0.664675294117647,0.7904858823529413,0.8974988235294117),
        rgb(0.2222222222222222 pt)=(0.6534188235294118,0.7835964705882353,0.8939913725490196),
        rgb(0.23232323232323232 pt)=(0.6365341176470588,0.7732623529411765,0.8887301960784313),
        rgb(0.24242424242424243 pt)=(0.6196494117647059,0.7629282352941177,0.8834690196078431),
        rgb(0.25252525252525254 pt)=(0.6083929411764706,0.7560388235294118,0.8799615686274509),
        rgb(0.26262626262626265 pt)=(0.5915082352941177,0.7457047058823529,0.8747003921568627),
        rgb(0.2727272727272727 pt)=(0.5802517647058825,0.7388152941176471,0.8711929411764705),
        rgb(0.2828282828282828 pt)=(0.5633670588235294,0.7284811764705883,0.8659317647058823),
        rgb(0.29292929292929293 pt)=(0.5521105882352941,0.7215917647058824,0.8624243137254901),
        rgb(0.30303030303030304 pt)=(0.5352258823529412,0.7112576470588235,0.8571631372549019),
        rgb(0.31313131313131315 pt)=(0.5183411764705883,0.7009235294117647,0.8519019607843137),
        rgb(0.32323232323232326 pt)=(0.507084705882353,0.6940341176470588,0.8483945098039215),
        rgb(0.3333333333333333 pt)=(0.4902,0.6837,0.8431333333333333),
        rgb(0.3434343434343434 pt)=(0.4789435294117647,0.6768105882352942,0.8396258823529411),
        rgb(0.35353535353535354 pt)=(0.46205882352941186,0.6664764705882353,0.8343647058823529),
        rgb(0.36363636363636365 pt)=(0.44517411764705883,0.6561423529411765,0.8291035294117647),
        rgb(0.37373737373737376 pt)=(0.4339176470588235,0.6492529411764706,0.8255960784313725),
        rgb(0.3838383838383838 pt)=(0.4170329411764706,0.6389188235294119,0.8203349019607843),
        rgb(0.3939393939393939 pt)=(0.4057764705882353,0.6320294117647058,0.8168274509803921),
        rgb(0.40404040404040403 pt)=(0.3888917647058824,0.621695294117647,0.8115662745098039),
        rgb(0.41414141414141414 pt)=(0.37200705882352947,0.6113611764705883,0.8063050980392157),
        rgb(0.42424242424242425 pt)=(0.36075058823529416,0.6044717647058824,0.8027976470588235),
        rgb(0.43434343434343436 pt)=(0.34386588235294113,0.5941376470588235,0.7975364705882353),
        rgb(0.4444444444444444 pt)=(0.33260941176470593,0.5872482352941177,0.7940290196078431),
        rgb(0.45454545454545453 pt)=(0.3157247058823529,0.5769141176470589,0.7887678431372549),
        rgb(0.46464646464646464 pt)=(0.3044682352941176,0.5700247058823529,0.7852603921568627),
        rgb(0.47474747474747475 pt)=(0.2875835294117647,0.5596905882352942,0.7799992156862745),
        rgb(0.48484848484848486 pt)=(0.27069882352941177,0.5493564705882352,0.7747380392156862),
        rgb(0.494949494949495 pt)=(0.25944235294117646,0.5424670588235294,0.7712305882352941),
        rgb(0.5050505050505051 pt)=(0.25524470588235293,0.5329164705882353,0.7612647058823528),
        rgb(0.5151515151515151 pt)=(0.2609043137254902,0.5270717647058824,0.7514843137254902),
        rgb(0.5252525252525253 pt)=(0.2693937254901961,0.5183047058823529,0.7368137254901961),
        rgb(0.5353535353535354 pt)=(0.277883137254902,0.5095376470588235,0.7221431372549019),
        rgb(0.5454545454545454 pt)=(0.28354274509803923,0.5036929411764706,0.7123627450980392),
        rgb(0.5555555555555556 pt)=(0.2920321568627451,0.49492588235294116,0.697692156862745),
        rgb(0.5656565656565656 pt)=(0.2976917647058824,0.48908117647058824,0.6879117647058823),
        rgb(0.5757575757575758 pt)=(0.30618117647058823,0.48031411764705884,0.6732411764705882),
        rgb(0.5858585858585859 pt)=(0.3118407843137255,0.47446941176470586,0.6634607843137255),
        rgb(0.5959595959595959 pt)=(0.3203301960784314,0.46570235294117646,0.6487901960784314),
        rgb(0.6060606060606061 pt)=(0.32881960784313724,0.45693529411764705,0.6341196078431373),
        rgb(0.6161616161616161 pt)=(0.33447921568627453,0.45109058823529413,0.6243392156862745),
        rgb(0.6262626262626263 pt)=(0.3429686274509804,0.4423235294117647,0.6096686274509804),
        rgb(0.6363636363636364 pt)=(0.3486282352941176,0.4364788235294118,0.5998882352941176),
        rgb(0.6464646464646465 pt)=(0.35711764705882354,0.42771176470588235,0.5852176470588235),
        rgb(0.6565656565656566 pt)=(0.3656070588235294,0.41894470588235294,0.5705470588235294),
        rgb(0.6666666666666666 pt)=(0.37126666666666663,0.4131,0.5607666666666666),
        rgb(0.6767676767676768 pt)=(0.37975607843137255,0.40433294117647056,0.5460960784313725),
        rgb(0.6868686868686869 pt)=(0.38541568627450984,0.39848823529411764,0.5363156862745098),
        rgb(0.696969696969697 pt)=(0.3939050980392157,0.38972117647058824,0.5216450980392157),
        rgb(0.7070707070707071 pt)=(0.40239450980392155,0.38095411764705883,0.5069745098039216),
        rgb(0.7171717171717171 pt)=(0.40805411764705884,0.3751094117647059,0.4971941176470588),
        rgb(0.7272727272727273 pt)=(0.4165435294117647,0.36634235294117645,0.4825235294117647),
        rgb(0.7373737373737373 pt)=(0.422203137254902,0.36049764705882353,0.47274313725490197),
        rgb(0.7474747474747475 pt)=(0.43069254901960785,0.35173058823529413,0.4580725490196078),
        rgb(0.7575757575757576 pt)=(0.4363521568627451,0.3458858823529412,0.4482921568627451),
        rgb(0.7676767676767676 pt)=(0.4448415686274509,0.33711882352941186,0.4336215686274511),
        rgb(0.7777777777777778 pt)=(0.4533309803921569,0.32835176470588234,0.41895098039215684),
        rgb(0.7878787878787878 pt)=(0.4589905882352941,0.3225070588235294,0.4091705882352941),
        rgb(0.797979797979798 pt)=(0.46748,0.31374,0.3945),
        rgb(0.8080808080808081 pt)=(0.47313960784313724,0.30789529411764704,0.38471960784313725),
        rgb(0.8181818181818182 pt)=(0.4816290196078431,0.2991282352941177,0.37004901960784314),
        rgb(0.8282828282828283 pt)=(0.49011843137254896,0.29036117647058834,0.35537843137254915),
        rgb(0.8383838383838383 pt)=(0.4957780392156863,0.2845164705882353,0.34559803921568627),
        rgb(0.8484848484848485 pt)=(0.5042674509803922,0.2757494117647059,0.33092745098039217),
        rgb(0.8585858585858586 pt)=(0.5099270588235294,0.26990470588235294,0.3211470588235294),
        rgb(0.8686868686868687 pt)=(0.5184164705882353,0.26113764705882353,0.30647647058823524),
        rgb(0.8787878787878788 pt)=(0.5240760784313725,0.2552929411764706,0.2966960784313726),
        rgb(0.8888888888888888 pt)=(0.5325654901960785,0.2465258823529412,0.2820254901960784),
        rgb(0.898989898989899 pt)=(0.5410549019607843,0.23775882352941174,0.26735490196078426),
        rgb(0.9090909090909091 pt)=(0.5467145098039216,0.23191411764705883,0.2575745098039216),
        rgb(0.9191919191919192 pt)=(0.5552039215686275,0.22314705882352942,0.2429039215686275),
        rgb(0.9292929292929293 pt)=(0.5608635294117648,0.2173023529411765,0.23312352941176473),
        rgb(0.9393939393939394 pt)=(0.5693529411764706,0.2085352941176471,0.21845294117647063),
        rgb(0.9494949494949495 pt)=(0.5778423529411765,0.19976823529411764,0.20378235294117653),
        rgb(0.9595959595959596 pt)=(0.5835019607843137,0.19392352941176472,0.19400196078431375),
        rgb(0.9696969696969697 pt)=(0.5919913725490196,0.1851564705882353,0.17933137254901965),
        rgb(0.9797979797979798 pt)=(0.5976509803921568,0.17931176470588234,0.16955098039215688),
        rgb(0.98989898989899 pt)=(0.6061403921568627,0.170544705882353,0.15488039215686278),
        rgb(1.0 pt)=(0.6118,0.1647,0.1451),},
    ]
    \addplot graphics [
        xmin = 0.000000,
        xmax = 361.000000,
        ymin = 0.000000,
        ymax = 70.000000
    ] {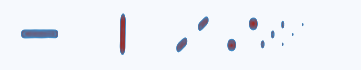};
\end{axis}
\end{tikzpicture}

%% file: tikz/OMP_reco_MUSE_axis0_largerScenario.tex
\begin{tikzpicture}
\begin{axis}[
       enlargelimits = false,
    axis on top = true,
    axis equal image,
    width=0.9\linewidth,
    ylabel = {\small mm},
    xtick = {0, 180, 361},
    xticklabels = {},
    ytick = {0,25,50},
    yticklabels = {0, 12.5, 25},
    colormap={FRI_white_blue_red}{
        rgb(0.0 pt)=(0.9686,0.9765,0.9922),
        rgb(0.010101010101010102 pt)=(0.9573435294117647,0.9696105882352941,0.9886925490196078),
        rgb(0.020202020202020204 pt)=(0.9404588235294118,0.9592764705882353,0.9834313725490196),
        rgb(0.030303030303030304 pt)=(0.9292023529411765,0.9523870588235295,0.9799239215686274),
        rgb(0.04040404040404041 pt)=(0.9123176470588236,0.9420529411764706,0.9746627450980392),
        rgb(0.050505050505050504 pt)=(0.9010611764705883,0.9351635294117647,0.971155294117647),
        rgb(0.06060606060606061 pt)=(0.8841764705882353,0.9248294117647059,0.9658941176470588),
        rgb(0.0707070707070707 pt)=(0.8672917647058824,0.9144952941176471,0.9606329411764706),
        rgb(0.08080808080808081 pt)=(0.8560352941176471,0.9076058823529412,0.9571254901960784),
        rgb(0.09090909090909091 pt)=(0.8391505882352941,0.8972717647058823,0.9518643137254902),
        rgb(0.10101010101010101 pt)=(0.8278941176470589,0.8903823529411765,0.948356862745098),
        rgb(0.1111111111111111 pt)=(0.8110094117647059,0.8800482352941177,0.9430956862745098),
        rgb(0.12121212121212122 pt)=(0.794124705882353,0.8697141176470589,0.9378345098039216),
        rgb(0.13131313131313133 pt)=(0.7828682352941176,0.8628247058823529,0.9343270588235294),
        rgb(0.1414141414141414 pt)=(0.7659835294117647,0.8524905882352941,0.9290658823529412),
        rgb(0.15151515151515152 pt)=(0.7547270588235294,0.8456011764705883,0.925558431372549),
        rgb(0.16161616161616163 pt)=(0.7378423529411765,0.8352670588235295,0.9202972549019608),
        rgb(0.1717171717171717 pt)=(0.7265858823529412,0.8283776470588236,0.9167898039215686),
        rgb(0.18181818181818182 pt)=(0.7097011764705883,0.8180435294117647,0.9115286274509804),
        rgb(0.1919191919191919 pt)=(0.6928164705882354,0.8077094117647059,0.9062674509803921),
        rgb(0.20202020202020202 pt)=(0.6815599999999999,0.80082,0.90276),
        rgb(0.21212121212121213 pt)=(0.664675294117647,0.7904858823529413,0.8974988235294117),
        rgb(0.2222222222222222 pt)=(0.6534188235294118,0.7835964705882353,0.8939913725490196),
        rgb(0.23232323232323232 pt)=(0.6365341176470588,0.7732623529411765,0.8887301960784313),
        rgb(0.24242424242424243 pt)=(0.6196494117647059,0.7629282352941177,0.8834690196078431),
        rgb(0.25252525252525254 pt)=(0.6083929411764706,0.7560388235294118,0.8799615686274509),
        rgb(0.26262626262626265 pt)=(0.5915082352941177,0.7457047058823529,0.8747003921568627),
        rgb(0.2727272727272727 pt)=(0.5802517647058825,0.7388152941176471,0.8711929411764705),
        rgb(0.2828282828282828 pt)=(0.5633670588235294,0.7284811764705883,0.8659317647058823),
        rgb(0.29292929292929293 pt)=(0.5521105882352941,0.7215917647058824,0.8624243137254901),
        rgb(0.30303030303030304 pt)=(0.5352258823529412,0.7112576470588235,0.8571631372549019),
        rgb(0.31313131313131315 pt)=(0.5183411764705883,0.7009235294117647,0.8519019607843137),
        rgb(0.32323232323232326 pt)=(0.507084705882353,0.6940341176470588,0.8483945098039215),
        rgb(0.3333333333333333 pt)=(0.4902,0.6837,0.8431333333333333),
        rgb(0.3434343434343434 pt)=(0.4789435294117647,0.6768105882352942,0.8396258823529411),
        rgb(0.35353535353535354 pt)=(0.46205882352941186,0.6664764705882353,0.8343647058823529),
        rgb(0.36363636363636365 pt)=(0.44517411764705883,0.6561423529411765,0.8291035294117647),
        rgb(0.37373737373737376 pt)=(0.4339176470588235,0.6492529411764706,0.8255960784313725),
        rgb(0.3838383838383838 pt)=(0.4170329411764706,0.6389188235294119,0.8203349019607843),
        rgb(0.3939393939393939 pt)=(0.4057764705882353,0.6320294117647058,0.8168274509803921),
        rgb(0.40404040404040403 pt)=(0.3888917647058824,0.621695294117647,0.8115662745098039),
        rgb(0.41414141414141414 pt)=(0.37200705882352947,0.6113611764705883,0.8063050980392157),
        rgb(0.42424242424242425 pt)=(0.36075058823529416,0.6044717647058824,0.8027976470588235),
        rgb(0.43434343434343436 pt)=(0.34386588235294113,0.5941376470588235,0.7975364705882353),
        rgb(0.4444444444444444 pt)=(0.33260941176470593,0.5872482352941177,0.7940290196078431),
        rgb(0.45454545454545453 pt)=(0.3157247058823529,0.5769141176470589,0.7887678431372549),
        rgb(0.46464646464646464 pt)=(0.3044682352941176,0.5700247058823529,0.7852603921568627),
        rgb(0.47474747474747475 pt)=(0.2875835294117647,0.5596905882352942,0.7799992156862745),
        rgb(0.48484848484848486 pt)=(0.27069882352941177,0.5493564705882352,0.7747380392156862),
        rgb(0.494949494949495 pt)=(0.25944235294117646,0.5424670588235294,0.7712305882352941),
        rgb(0.5050505050505051 pt)=(0.25524470588235293,0.5329164705882353,0.7612647058823528),
        rgb(0.5151515151515151 pt)=(0.2609043137254902,0.5270717647058824,0.7514843137254902),
        rgb(0.5252525252525253 pt)=(0.2693937254901961,0.5183047058823529,0.7368137254901961),
        rgb(0.5353535353535354 pt)=(0.277883137254902,0.5095376470588235,0.7221431372549019),
        rgb(0.5454545454545454 pt)=(0.28354274509803923,0.5036929411764706,0.7123627450980392),
        rgb(0.5555555555555556 pt)=(0.2920321568627451,0.49492588235294116,0.697692156862745),
        rgb(0.5656565656565656 pt)=(0.2976917647058824,0.48908117647058824,0.6879117647058823),
        rgb(0.5757575757575758 pt)=(0.30618117647058823,0.48031411764705884,0.6732411764705882),
        rgb(0.5858585858585859 pt)=(0.3118407843137255,0.47446941176470586,0.6634607843137255),
        rgb(0.5959595959595959 pt)=(0.3203301960784314,0.46570235294117646,0.6487901960784314),
        rgb(0.6060606060606061 pt)=(0.32881960784313724,0.45693529411764705,0.6341196078431373),
        rgb(0.6161616161616161 pt)=(0.33447921568627453,0.45109058823529413,0.6243392156862745),
        rgb(0.6262626262626263 pt)=(0.3429686274509804,0.4423235294117647,0.6096686274509804),
        rgb(0.6363636363636364 pt)=(0.3486282352941176,0.4364788235294118,0.5998882352941176),
        rgb(0.6464646464646465 pt)=(0.35711764705882354,0.42771176470588235,0.5852176470588235),
        rgb(0.6565656565656566 pt)=(0.3656070588235294,0.41894470588235294,0.5705470588235294),
        rgb(0.6666666666666666 pt)=(0.37126666666666663,0.4131,0.5607666666666666),
        rgb(0.6767676767676768 pt)=(0.37975607843137255,0.40433294117647056,0.5460960784313725),
        rgb(0.6868686868686869 pt)=(0.38541568627450984,0.39848823529411764,0.5363156862745098),
        rgb(0.696969696969697 pt)=(0.3939050980392157,0.38972117647058824,0.5216450980392157),
        rgb(0.7070707070707071 pt)=(0.40239450980392155,0.38095411764705883,0.5069745098039216),
        rgb(0.7171717171717171 pt)=(0.40805411764705884,0.3751094117647059,0.4971941176470588),
        rgb(0.7272727272727273 pt)=(0.4165435294117647,0.36634235294117645,0.4825235294117647),
        rgb(0.7373737373737373 pt)=(0.422203137254902,0.36049764705882353,0.47274313725490197),
        rgb(0.7474747474747475 pt)=(0.43069254901960785,0.35173058823529413,0.4580725490196078),
        rgb(0.7575757575757576 pt)=(0.4363521568627451,0.3458858823529412,0.4482921568627451),
        rgb(0.7676767676767676 pt)=(0.4448415686274509,0.33711882352941186,0.4336215686274511),
        rgb(0.7777777777777778 pt)=(0.4533309803921569,0.32835176470588234,0.41895098039215684),
        rgb(0.7878787878787878 pt)=(0.4589905882352941,0.3225070588235294,0.4091705882352941),
        rgb(0.797979797979798 pt)=(0.46748,0.31374,0.3945),
        rgb(0.8080808080808081 pt)=(0.47313960784313724,0.30789529411764704,0.38471960784313725),
        rgb(0.8181818181818182 pt)=(0.4816290196078431,0.2991282352941177,0.37004901960784314),
        rgb(0.8282828282828283 pt)=(0.49011843137254896,0.29036117647058834,0.35537843137254915),
        rgb(0.8383838383838383 pt)=(0.4957780392156863,0.2845164705882353,0.34559803921568627),
        rgb(0.8484848484848485 pt)=(0.5042674509803922,0.2757494117647059,0.33092745098039217),
        rgb(0.8585858585858586 pt)=(0.5099270588235294,0.26990470588235294,0.3211470588235294),
        rgb(0.8686868686868687 pt)=(0.5184164705882353,0.26113764705882353,0.30647647058823524),
        rgb(0.8787878787878788 pt)=(0.5240760784313725,0.2552929411764706,0.2966960784313726),
        rgb(0.8888888888888888 pt)=(0.5325654901960785,0.2465258823529412,0.2820254901960784),
        rgb(0.898989898989899 pt)=(0.5410549019607843,0.23775882352941174,0.26735490196078426),
        rgb(0.9090909090909091 pt)=(0.5467145098039216,0.23191411764705883,0.2575745098039216),
        rgb(0.9191919191919192 pt)=(0.5552039215686275,0.22314705882352942,0.2429039215686275),
        rgb(0.9292929292929293 pt)=(0.5608635294117648,0.2173023529411765,0.23312352941176473),
        rgb(0.9393939393939394 pt)=(0.5693529411764706,0.2085352941176471,0.21845294117647063),
        rgb(0.9494949494949495 pt)=(0.5778423529411765,0.19976823529411764,0.20378235294117653),
        rgb(0.9595959595959596 pt)=(0.5835019607843137,0.19392352941176472,0.19400196078431375),
        rgb(0.9696969696969697 pt)=(0.5919913725490196,0.1851564705882353,0.17933137254901965),
        rgb(0.9797979797979798 pt)=(0.5976509803921568,0.17931176470588234,0.16955098039215688),
        rgb(0.98989898989899 pt)=(0.6061403921568627,0.170544705882353,0.15488039215686278),
        rgb(1.0 pt)=(0.6118,0.1647,0.1451),},
    ]
    \addplot graphics [
        xmin = 0.000000,
        xmax = 361.000000,
        ymin = 0.000000,
        ymax = 70.000000
    ] {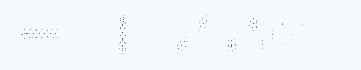};
\end{axis}
\end{tikzpicture}

%% file: tikz/SAFT_reco_MUSE_axis0_largerScenario.tex
\begin{tikzpicture}
\begin{axis}[
    enlargelimits = false,
    axis on top = true,
    axis equal image,
    width=0.9\linewidth,
    xlabel shift = -4pt,
    xlabel = {\small mm},
    ylabel = {\small mm},
    xtick = {0,180,361},
    xticklabels = {0,90, 180.5},
    ytick = {0,25,50},
    yticklabels = {0, 12.5, 25},
    colormap={FRI_white_blue_red}{
        rgb(0.0 pt)=(0.9686,0.9765,0.9922),
        rgb(0.010101010101010102 pt)=(0.9573435294117647,0.9696105882352941,0.9886925490196078),
        rgb(0.020202020202020204 pt)=(0.9404588235294118,0.9592764705882353,0.9834313725490196),
        rgb(0.030303030303030304 pt)=(0.9292023529411765,0.9523870588235295,0.9799239215686274),
        rgb(0.04040404040404041 pt)=(0.9123176470588236,0.9420529411764706,0.9746627450980392),
        rgb(0.050505050505050504 pt)=(0.9010611764705883,0.9351635294117647,0.971155294117647),
        rgb(0.06060606060606061 pt)=(0.8841764705882353,0.9248294117647059,0.9658941176470588),
        rgb(0.0707070707070707 pt)=(0.8672917647058824,0.9144952941176471,0.9606329411764706),
        rgb(0.08080808080808081 pt)=(0.8560352941176471,0.9076058823529412,0.9571254901960784),
        rgb(0.09090909090909091 pt)=(0.8391505882352941,0.8972717647058823,0.9518643137254902),
        rgb(0.10101010101010101 pt)=(0.8278941176470589,0.8903823529411765,0.948356862745098),
        rgb(0.1111111111111111 pt)=(0.8110094117647059,0.8800482352941177,0.9430956862745098),
        rgb(0.12121212121212122 pt)=(0.794124705882353,0.8697141176470589,0.9378345098039216),
        rgb(0.13131313131313133 pt)=(0.7828682352941176,0.8628247058823529,0.9343270588235294),
        rgb(0.1414141414141414 pt)=(0.7659835294117647,0.8524905882352941,0.9290658823529412),
        rgb(0.15151515151515152 pt)=(0.7547270588235294,0.8456011764705883,0.925558431372549),
        rgb(0.16161616161616163 pt)=(0.7378423529411765,0.8352670588235295,0.9202972549019608),
        rgb(0.1717171717171717 pt)=(0.7265858823529412,0.8283776470588236,0.9167898039215686),
        rgb(0.18181818181818182 pt)=(0.7097011764705883,0.8180435294117647,0.9115286274509804),
        rgb(0.1919191919191919 pt)=(0.6928164705882354,0.8077094117647059,0.9062674509803921),
        rgb(0.20202020202020202 pt)=(0.6815599999999999,0.80082,0.90276),
        rgb(0.21212121212121213 pt)=(0.664675294117647,0.7904858823529413,0.8974988235294117),
        rgb(0.2222222222222222 pt)=(0.6534188235294118,0.7835964705882353,0.8939913725490196),
        rgb(0.23232323232323232 pt)=(0.6365341176470588,0.7732623529411765,0.8887301960784313),
        rgb(0.24242424242424243 pt)=(0.6196494117647059,0.7629282352941177,0.8834690196078431),
        rgb(0.25252525252525254 pt)=(0.6083929411764706,0.7560388235294118,0.8799615686274509),
        rgb(0.26262626262626265 pt)=(0.5915082352941177,0.7457047058823529,0.8747003921568627),
        rgb(0.2727272727272727 pt)=(0.5802517647058825,0.7388152941176471,0.8711929411764705),
        rgb(0.2828282828282828 pt)=(0.5633670588235294,0.7284811764705883,0.8659317647058823),
        rgb(0.29292929292929293 pt)=(0.5521105882352941,0.7215917647058824,0.8624243137254901),
        rgb(0.30303030303030304 pt)=(0.5352258823529412,0.7112576470588235,0.8571631372549019),
        rgb(0.31313131313131315 pt)=(0.5183411764705883,0.7009235294117647,0.8519019607843137),
        rgb(0.32323232323232326 pt)=(0.507084705882353,0.6940341176470588,0.8483945098039215),
        rgb(0.3333333333333333 pt)=(0.4902,0.6837,0.8431333333333333),
        rgb(0.3434343434343434 pt)=(0.4789435294117647,0.6768105882352942,0.8396258823529411),
        rgb(0.35353535353535354 pt)=(0.46205882352941186,0.6664764705882353,0.8343647058823529),
        rgb(0.36363636363636365 pt)=(0.44517411764705883,0.6561423529411765,0.8291035294117647),
        rgb(0.37373737373737376 pt)=(0.4339176470588235,0.6492529411764706,0.8255960784313725),
        rgb(0.3838383838383838 pt)=(0.4170329411764706,0.6389188235294119,0.8203349019607843),
        rgb(0.3939393939393939 pt)=(0.4057764705882353,0.6320294117647058,0.8168274509803921),
        rgb(0.40404040404040403 pt)=(0.3888917647058824,0.621695294117647,0.8115662745098039),
        rgb(0.41414141414141414 pt)=(0.37200705882352947,0.6113611764705883,0.8063050980392157),
        rgb(0.42424242424242425 pt)=(0.36075058823529416,0.6044717647058824,0.8027976470588235),
        rgb(0.43434343434343436 pt)=(0.34386588235294113,0.5941376470588235,0.7975364705882353),
        rgb(0.4444444444444444 pt)=(0.33260941176470593,0.5872482352941177,0.7940290196078431),
        rgb(0.45454545454545453 pt)=(0.3157247058823529,0.5769141176470589,0.7887678431372549),
        rgb(0.46464646464646464 pt)=(0.3044682352941176,0.5700247058823529,0.7852603921568627),
        rgb(0.47474747474747475 pt)=(0.2875835294117647,0.5596905882352942,0.7799992156862745),
        rgb(0.48484848484848486 pt)=(0.27069882352941177,0.5493564705882352,0.7747380392156862),
        rgb(0.494949494949495 pt)=(0.25944235294117646,0.5424670588235294,0.7712305882352941),
        rgb(0.5050505050505051 pt)=(0.25524470588235293,0.5329164705882353,0.7612647058823528),
        rgb(0.5151515151515151 pt)=(0.2609043137254902,0.5270717647058824,0.7514843137254902),
        rgb(0.5252525252525253 pt)=(0.2693937254901961,0.5183047058823529,0.7368137254901961),
        rgb(0.5353535353535354 pt)=(0.277883137254902,0.5095376470588235,0.7221431372549019),
        rgb(0.5454545454545454 pt)=(0.28354274509803923,0.5036929411764706,0.7123627450980392),
        rgb(0.5555555555555556 pt)=(0.2920321568627451,0.49492588235294116,0.697692156862745),
        rgb(0.5656565656565656 pt)=(0.2976917647058824,0.48908117647058824,0.6879117647058823),
        rgb(0.5757575757575758 pt)=(0.30618117647058823,0.48031411764705884,0.6732411764705882),
        rgb(0.5858585858585859 pt)=(0.3118407843137255,0.47446941176470586,0.6634607843137255),
        rgb(0.5959595959595959 pt)=(0.3203301960784314,0.46570235294117646,0.6487901960784314),
        rgb(0.6060606060606061 pt)=(0.32881960784313724,0.45693529411764705,0.6341196078431373),
        rgb(0.6161616161616161 pt)=(0.33447921568627453,0.45109058823529413,0.6243392156862745),
        rgb(0.6262626262626263 pt)=(0.3429686274509804,0.4423235294117647,0.6096686274509804),
        rgb(0.6363636363636364 pt)=(0.3486282352941176,0.4364788235294118,0.5998882352941176),
        rgb(0.6464646464646465 pt)=(0.35711764705882354,0.42771176470588235,0.5852176470588235),
        rgb(0.6565656565656566 pt)=(0.3656070588235294,0.41894470588235294,0.5705470588235294),
        rgb(0.6666666666666666 pt)=(0.37126666666666663,0.4131,0.5607666666666666),
        rgb(0.6767676767676768 pt)=(0.37975607843137255,0.40433294117647056,0.5460960784313725),
        rgb(0.6868686868686869 pt)=(0.38541568627450984,0.39848823529411764,0.5363156862745098),
        rgb(0.696969696969697 pt)=(0.3939050980392157,0.38972117647058824,0.5216450980392157),
        rgb(0.7070707070707071 pt)=(0.40239450980392155,0.38095411764705883,0.5069745098039216),
        rgb(0.7171717171717171 pt)=(0.40805411764705884,0.3751094117647059,0.4971941176470588),
        rgb(0.7272727272727273 pt)=(0.4165435294117647,0.36634235294117645,0.4825235294117647),
        rgb(0.7373737373737373 pt)=(0.422203137254902,0.36049764705882353,0.47274313725490197),
        rgb(0.7474747474747475 pt)=(0.43069254901960785,0.35173058823529413,0.4580725490196078),
        rgb(0.7575757575757576 pt)=(0.4363521568627451,0.3458858823529412,0.4482921568627451),
        rgb(0.7676767676767676 pt)=(0.4448415686274509,0.33711882352941186,0.4336215686274511),
        rgb(0.7777777777777778 pt)=(0.4533309803921569,0.32835176470588234,0.41895098039215684),
        rgb(0.7878787878787878 pt)=(0.4589905882352941,0.3225070588235294,0.4091705882352941),
        rgb(0.797979797979798 pt)=(0.46748,0.31374,0.3945),
        rgb(0.8080808080808081 pt)=(0.47313960784313724,0.30789529411764704,0.38471960784313725),
        rgb(0.8181818181818182 pt)=(0.4816290196078431,0.2991282352941177,0.37004901960784314),
        rgb(0.8282828282828283 pt)=(0.49011843137254896,0.29036117647058834,0.35537843137254915),
        rgb(0.8383838383838383 pt)=(0.4957780392156863,0.2845164705882353,0.34559803921568627),
        rgb(0.8484848484848485 pt)=(0.5042674509803922,0.2757494117647059,0.33092745098039217),
        rgb(0.8585858585858586 pt)=(0.5099270588235294,0.26990470588235294,0.3211470588235294),
        rgb(0.8686868686868687 pt)=(0.5184164705882353,0.26113764705882353,0.30647647058823524),
        rgb(0.8787878787878788 pt)=(0.5240760784313725,0.2552929411764706,0.2966960784313726),
        rgb(0.8888888888888888 pt)=(0.5325654901960785,0.2465258823529412,0.2820254901960784),
        rgb(0.898989898989899 pt)=(0.5410549019607843,0.23775882352941174,0.26735490196078426),
        rgb(0.9090909090909091 pt)=(0.5467145098039216,0.23191411764705883,0.2575745098039216),
        rgb(0.9191919191919192 pt)=(0.5552039215686275,0.22314705882352942,0.2429039215686275),
        rgb(0.9292929292929293 pt)=(0.5608635294117648,0.2173023529411765,0.23312352941176473),
        rgb(0.9393939393939394 pt)=(0.5693529411764706,0.2085352941176471,0.21845294117647063),
        rgb(0.9494949494949495 pt)=(0.5778423529411765,0.19976823529411764,0.20378235294117653),
        rgb(0.9595959595959596 pt)=(0.5835019607843137,0.19392352941176472,0.19400196078431375),
        rgb(0.9696969696969697 pt)=(0.5919913725490196,0.1851564705882353,0.17933137254901965),
        rgb(0.9797979797979798 pt)=(0.5976509803921568,0.17931176470588234,0.16955098039215688),
        rgb(0.98989898989899 pt)=(0.6061403921568627,0.170544705882353,0.15488039215686278),
        rgb(1.0 pt)=(0.6118,0.1647,0.1451),},
    ]
    \addplot graphics [
        xmin = 0.000000,
        xmax = 361.000000,
        ymin = 0.000000,
        ymax = 70.000000
    ] {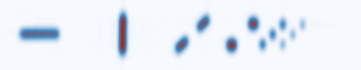};
\end{axis}
\end{tikzpicture}

%% file: tikz/muse_global_random_dm_df_single_coefficient.tex
\begin{tikzpicture}

\begin{groupplot}[group style={group size=1 by 2, vertical sep=0, x descriptions at=edge bottom}]
\nextgroupplot[
axis equal image,
axis on top=true,
height= ,
scaled x ticks=manual:{}{\pgfmathparse{#1}},
tick align=outside,
tick pos=left,
width=\textwidth,
x grid style={white!69.0196078431373!black},
xmin=0, xmax=190.5,
xtick style={color=black},
xticklabels={},
y grid style={white!69.0196078431373!black},
ylabel={\(\displaystyle y\) in mm},
ymin=0, ymax=35,
ytick style={color=black},
ytick={5,25},
yticklabels={\(\displaystyle 5\),\(\displaystyle 25\)}
]
\addplot graphics [includegraphics cmd=\pgfimage,xmin=0, xmax=190.5, ymin=0, ymax=35] {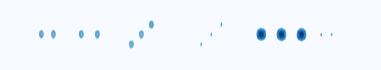};
\input{tikz/muse_ground_truth_top_view.tex}

\nextgroupplot[
axis equal image,
axis on top=true,
height= ,
tick align=outside,
tick pos=left,
width=\textwidth,
x grid style={white!69.0196078431373!black},
xlabel={\(\displaystyle x\) in mm},
xmin=0, xmax=190.5,
xtick style={color=black},
xtick={0,25,50,75,100,125,150,175,200},
xticklabels={\(\displaystyle 0\),\(\displaystyle 25\),\(\displaystyle 50\),\(\displaystyle 75\),\(\displaystyle 100\),\(\displaystyle 125\),\(\displaystyle 150\),\(\displaystyle 175\),\(\displaystyle 200\)},
y dir=reverse,
y grid style={white!69.0196078431373!black},
ylabel={\(\displaystyle z\) in mm},
ymin=67.1125, ymax=89.2375,
ytick style={color=black},
ytick={75,85},
yticklabels={\(\displaystyle 75\),\(\displaystyle 85\)}
]
\addplot graphics [includegraphics cmd=\pgfimage,xmin=0, xmax=190.5, ymin=67.1125, ymax=89.2375] {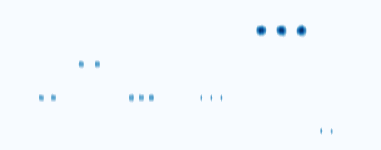};
\input{tikz/muse_ground_truth_side_view.tex}
\end{groupplot}
\end{tikzpicture}

%% file: tikz/muse_ground_truth_side_view.tex
\newcommand{\yPosOne}{18}
\newcommand{\yPosTwo}{13}
\newcommand{\circleRadiusOne}{1.5}
\newcommand{\circleRadiusTwo}{0.75}
\newcommand{\circleRadiusThree}{2.75}
\newcommand{\offsetOne}{20.7}
\newcommand{\offsetTwo}{65.7}
\newcommand{\offsetThree}{100.7}
\newcommand{\offsetFour}{160.7}
\newcommand{\offsetFive}{130.7}
\newcommand{\TenMMDeepHoles}{81.5}
\newcommand{\FifteenMMDeepHoles}{76.5}
\newcommand{\TwentyMMDeepHoles}{71.5}
\newcommand{\FiveMMDeepHoles}{86.5}
\newcommand{\linestyleGT}{densely dotted}
\newcommand{\linestyleFBHWalls}{dashed}

\draw[\linestyleGT] (\offsetOne-\circleRadiusOne, \TenMMDeepHoles) -- (\offsetOne+\circleRadiusOne, \TenMMDeepHoles) {};
\draw[\linestyleFBHWalls] (\offsetOne-\circleRadiusOne, 90) -- (\offsetOne-\circleRadiusOne, \TenMMDeepHoles){};
\draw[\linestyleFBHWalls] (\offsetOne+\circleRadiusOne, 90) -- (\offsetOne+\circleRadiusOne, \TenMMDeepHoles){};
\draw[\linestyleGT] (\offsetOne+6-\circleRadiusOne, \TenMMDeepHoles) -- (\offsetOne+6+\circleRadiusOne, \TenMMDeepHoles) {};
\draw[\linestyleFBHWalls] (\offsetOne+6-\circleRadiusOne, 90) -- (\offsetOne+6-\circleRadiusOne, \TenMMDeepHoles){};
\draw[\linestyleFBHWalls] (\offsetOne+6+\circleRadiusOne, 90) -- (\offsetOne+6+\circleRadiusOne, \TenMMDeepHoles){};
\draw[\linestyleGT] (\offsetOne+20-\circleRadiusOne, \FifteenMMDeepHoles) -- (\offsetOne+20+\circleRadiusOne, \FifteenMMDeepHoles) {};
\draw[\linestyleFBHWalls] (\offsetOne+20-\circleRadiusOne, 90) -- (\offsetOne+20-\circleRadiusOne, \FifteenMMDeepHoles){};
\draw[\linestyleFBHWalls] (\offsetOne+20+\circleRadiusOne, 90) -- (\offsetOne+20+\circleRadiusOne, \FifteenMMDeepHoles){};
\draw[\linestyleGT] (\offsetOne+28-\circleRadiusOne, \FifteenMMDeepHoles) -- (\offsetOne+28+\circleRadiusOne, \FifteenMMDeepHoles) {};
\draw[\linestyleFBHWalls] (\offsetOne+28-\circleRadiusOne, 90) -- (\offsetOne+28-\circleRadiusOne, \FifteenMMDeepHoles){};
\draw[\linestyleFBHWalls] (\offsetOne+28+\circleRadiusOne, 90) -- (\offsetOne+28+\circleRadiusOne, \FifteenMMDeepHoles){};

\draw[\linestyleGT] (\offsetTwo-\circleRadiusOne, \TenMMDeepHoles) -- (\offsetTwo+\circleRadiusOne, \TenMMDeepHoles) {};
\draw[\linestyleFBHWalls] (\offsetTwo-\circleRadiusOne, 90) -- (\offsetTwo-\circleRadiusOne, \TenMMDeepHoles) {};
\draw[\linestyleFBHWalls] (\offsetTwo+\circleRadiusOne, 90) -- (\offsetTwo+\circleRadiusOne, \TenMMDeepHoles) {};
\draw[\linestyleGT] (\offsetTwo+5-\circleRadiusOne, \TenMMDeepHoles) -- (\offsetTwo+5+\circleRadiusOne, \TenMMDeepHoles) {};
\draw[\linestyleFBHWalls] (\offsetTwo+5-\circleRadiusOne, 90) -- (\offsetTwo+5-\circleRadiusOne, \TenMMDeepHoles) {};
\draw[\linestyleFBHWalls] (\offsetTwo+5+\circleRadiusOne, 90) -- (\offsetTwo+5+\circleRadiusOne, \TenMMDeepHoles) {};
\draw[\linestyleGT] (\offsetTwo+10-\circleRadiusOne, \TenMMDeepHoles) -- (\offsetTwo+10+\circleRadiusOne, \TenMMDeepHoles) {};
\draw[\linestyleFBHWalls] (\offsetTwo+10-\circleRadiusOne, 90) -- (\offsetTwo+10-\circleRadiusOne, \TenMMDeepHoles) {};
\draw[\linestyleFBHWalls] (\offsetTwo+10+\circleRadiusOne, 90) -- (\offsetTwo+10+\circleRadiusOne, \TenMMDeepHoles) {};

\draw[\linestyleGT] (\offsetThree-\circleRadiusTwo, \TenMMDeepHoles) -- (\offsetThree+\circleRadiusTwo, \TenMMDeepHoles) {};
\draw[\linestyleFBHWalls] (\offsetThree-\circleRadiusTwo, 90) -- (\offsetThree-\circleRadiusTwo, \TenMMDeepHoles) {};
\draw[\linestyleFBHWalls] (\offsetThree+\circleRadiusTwo, 90) -- (\offsetThree+\circleRadiusTwo, \TenMMDeepHoles) {};
\draw[\linestyleGT] (\offsetThree+5-\circleRadiusTwo, \TenMMDeepHoles) -- (\offsetThree+5+\circleRadiusTwo, \TenMMDeepHoles) {};
\draw[\linestyleFBHWalls] (\offsetThree+5-\circleRadiusTwo, 90) -- (\offsetThree+5-\circleRadiusTwo, \TenMMDeepHoles) {};
\draw[\linestyleFBHWalls] (\offsetThree+5+\circleRadiusTwo, 90) -- (\offsetThree+5+\circleRadiusTwo, \TenMMDeepHoles) {};
\draw[\linestyleGT] (\offsetThree+10-\circleRadiusTwo, \TenMMDeepHoles) -- (\offsetThree+10+\circleRadiusTwo, \TenMMDeepHoles) {};
\draw[\linestyleFBHWalls] (\offsetThree+10-\circleRadiusTwo, 90) -- (\offsetThree+10-\circleRadiusTwo, \TenMMDeepHoles) {};
\draw[\linestyleFBHWalls] (\offsetThree+10+\circleRadiusTwo, 90) -- (\offsetThree+10+\circleRadiusTwo, \TenMMDeepHoles) {};

\draw[\linestyleGT] (\offsetFive-\circleRadiusThree, \TwentyMMDeepHoles) -- (\offsetFive+\circleRadiusThree, \TwentyMMDeepHoles) {};
\draw[\linestyleFBHWalls] (\offsetFive-\circleRadiusThree, 90) --
(\offsetFive-\circleRadiusThree, \TwentyMMDeepHoles) {};
\draw[\linestyleFBHWalls] (\offsetFive+\circleRadiusThree, 90) --
(\offsetFive+\circleRadiusThree, \TwentyMMDeepHoles) {};
\draw[\linestyleGT] (\offsetFive+10-\circleRadiusThree, \TwentyMMDeepHoles) -- (\offsetFive+10+\circleRadiusThree, \TwentyMMDeepHoles) {};
\draw[\linestyleFBHWalls] (\offsetFive+10-\circleRadiusThree, 90) --
(\offsetFive+10-\circleRadiusThree, \TwentyMMDeepHoles) {};
\draw[\linestyleFBHWalls] (\offsetFive+10+\circleRadiusThree, 90) --
(\offsetFive+10+\circleRadiusThree, \TwentyMMDeepHoles) {};

\draw[\linestyleGT] (\offsetFive+20-\circleRadiusThree, \TwentyMMDeepHoles) -- (\offsetFive+20+\circleRadiusThree, \TwentyMMDeepHoles) {};
\draw[\linestyleFBHWalls] (\offsetFive+20-\circleRadiusThree, 90) --
(\offsetFive+20-\circleRadiusThree, \TwentyMMDeepHoles) {};
\draw[\linestyleFBHWalls] (\offsetFive+20+\circleRadiusThree, 90) --
(\offsetFive+20+\circleRadiusThree, \TwentyMMDeepHoles) {};

\draw[\linestyleGT] (\offsetFour-\circleRadiusTwo, \FiveMMDeepHoles) -- (\offsetFour+\circleRadiusTwo, \FiveMMDeepHoles) {};
\draw[\linestyleFBHWalls] (\offsetFour-\circleRadiusTwo,90) --(\offsetFour-\circleRadiusTwo, \FiveMMDeepHoles) {};
\draw[\linestyleFBHWalls] (\offsetFour+\circleRadiusTwo,90) --(\offsetFour+\circleRadiusTwo, \FiveMMDeepHoles) {};

\draw[\linestyleGT] (\offsetFour+5.25-\circleRadiusTwo, \FiveMMDeepHoles) -- (\offsetFour+5.25+\circleRadiusTwo, \FiveMMDeepHoles) {};
\draw[\linestyleFBHWalls] (\offsetFour+5.25-\circleRadiusTwo,90) --(\offsetFour+5.25-\circleRadiusTwo, \FiveMMDeepHoles) {};
\draw[\linestyleFBHWalls] (\offsetFour+5.25+\circleRadiusTwo,90) --(\offsetFour+5.25+\circleRadiusTwo, \FiveMMDeepHoles) {};

%% file: sections/impact.tex

%
%
%
%
%

As we have shown in \Cref{examples}, the presented package provides several major benefits for improving the efficiency of research tasks involving large-scale linear structured operators.
By design of the \gls{api}, presented in \Cref{description}, a research programmer can abstract low-level optimization away from high-level algorithm design without sacrificing general performance optimizations, resulting in greatly improved readability, extensibility and portability of the produced code.
Linear mappings become implementable very rapidly and due to the object-oriented approach resembling construction-kits, reuse potential of matrix-free complex linear transforms and algorithms is greatly elevated, especially by the abstraction model provided by this package.
Due to the included testing functionality it is possible to debug implementations at every abstraction level, greatly reducing the time required for debug efforts.
The consequent abstraction model allows for the separation of efficiency-boosting implementation tweaks at every level of the flexible model architecture, up to the algorithmic design.
This allows to find common silly mistakes early in the debug cycle and maintains clarity of code and its interfaces.
All those features result in \fastmat being a tool to boost the work efficiency of researchers or engineers that regularly have to deal with large-scale structural models and algorithms that apply those.

In the EMS \rev{at the Technische Universität Ilmenau} research group, the package fueled several research activities involving compressed sensing on very large structured linear signal models.
As a consequence of the packages focus on clarity, flexibility and reuse, it was possible to achieve higher research output compared to writing conventional research code.
Especially the ability to easily incorporate existing results in following projects and the possibility of adapting existing solutions to new related problems drastically reduced the required coding effort.
Following, a few examples shall indicate how and why users decided to utilize the package for their research code implementations

In ultrasonic non-destructive testing, the sophisticated model for 3D defect detection in specimens can be implemented almost exactly like written in the research publications.
Due to the insanely large problem sizes, implementing such models involved extensive amounts of programming and debugging.
The abstraction approach allowed to formulate complex models easier, resulting in cleaner and easier to follow implementations.
Many research teams are faced with a ``specialization trap'', where every colleague maintains a dedicated piece of the project that others do not dare trying to understand.
This puts a direct limit on the sustainable problem complexity and team efficiency -- especially when teams grow in size -- since communication between team members is limited (even already by available time) ~\cite{BrooThem1995}.
Our package is a solution to breaking this specialization trap by enabling researches to better understand complex problems and thus allowing them to collaborate more efficiently in even larger projects to solve more complex problems.

On top of the aspect of team dynamics, the resulting implementation efficiency is also sufficiently close to the optimum, where the gap in efficiency only stems from the generality of the implementation.
Therefore, the best of both worlds can be achieved: Convenience no longer needed to be sacrificed for speed (and vice-versa), leading to a drastic increase in project team interaction.
As a result this carved off several months of development time, which could be spent in improving the package.
The flexibility to modify and extend models without complicated code redesigns allowed to add a compressed sensing scheme with only minimal changes to the model and no modification at the side of the sparsity-based reconstruction algorithms.

Benefits not only occurred at the end of computation performance.
Especially in memory consumption the abstraction approach payed off by reducing the space required for dense model representations from several petabytes to less than 1 TB.
This drastic reduction made compressed-sensing based 3D reconstruction of defects from ultrasound measurements possible for the first time, which was previously deemed unfeasible due to the vast model sizes involved~\cite{journal-tuffc-2021,semper2018defectdetection,Krieg2018onlineSAFT,semper20193dfista,kirchhof2018gpuUS,Krieg2020pipes}.

In a different field, \fastmat powered the investigation into a new method of compressed-sensing ultra wideband radar~\cite{wagner_csuwb_2021}.
Here, parameter studies yielded the impact of certain model decisions to investigate the impact of circuit- and architecture-level design decisions for radar frontends.
Due to the construction-kit structure of this package, plug-and-play swapping of inner model abstraction layers was possible, allowing the very rapid investigation of different circuit architectures with extensive numerical support.
This is also beneficial for the design of new hybrid frontend architectures for \gls{gnss}.
The structure of the \gls{prn} signals at the heart of such localization concepts, and the fact that efficient implementations of convolution operators exist, invite the use of \fastmat for modeling multi-channel \gls{gnss} scenarios, too.
In particular, the built-in \texttt{LFSRCirculant} class applies efficient cyclic convolution with any \gls{prn} signal, that can be produced by an hardware-efficient \gls{lfsr}.
Similar to the ultrasound example, \fastmat's abstract capabilities can also be utilized to extend the scope of this work by adding practical validation techniques, such as \gls{ota} testing and 3D \gls{wfs}~\cite{Vintimilla2021RealisticEO}.

In the field of machine learning, our package is utilized to perform \gls{ski} for scaling \gls{gp} to massive datasets on the example of three-dimensional weather radar datasets with over 100 million points~\cite{pmlr-v130-yadav21a}.
One mentionable application of our algorithmic subsystem, being used for its high-performing efficient rank-1 update \gls{omp} implementation, is in pygpc, a sensitivity and uncertainty analysis toolbox for Python~\cite{WEISE2020100450}.

%% file: sections/conclusion.tex


We have shown that with a carefully designed \gls{api} one can design an architecture that unifies dense and matrix-free linear mappings, while leveraging the strengths of both approaches, and keeping the complications between them at a minimum.

As we demonstrate this results in simplifications both in terms of the written code as well as the design process that is needed to construct a specific implementation.

Future versions of the package will include support for tensor unfoldings in order to realize higher-order linear mappings more rigorously.
Further, improved support for the specification of more complex block structures aims to make better use of the redundancy occurring from repeating structures within a matrix-presentation.

In essence, the described package allows to focus more on the actual problems at hand, when dealing with large-scale structured linear models, without being subjected to the obstacles of realizing them from scratch.